\documentclass[aps,pre,notitlepage,twocolumn]{revtex4-1}
\usepackage{amssymb}
\usepackage{amsmath}
\usepackage{graphicx}
\usepackage{color}
\usepackage{subcaption}

\newcommand{\sgn}{\hbox{sgn}}
\newcommand{\bs}{\boldsymbol}

\begin{document}

\title{Soliton gas in bidirectional dispersive hydrodynamics}

\author{Thibault Congy}
\email[Corresponding author: ]{thibault.congy@northumbria.ac.uk}
\author{Gennady  El} 
\author{Giacomo Roberti}
\affiliation{Department of Mathematics, Physics and Electrical
Engineering, Northumbria University, Newcastle upon Tyne, United
Kingdom}

\begin{abstract}
The  theory of soliton gas had been previously developed for
unidirectional integrable dispersive hydrodynamics in which the
soliton gas properties are determined by the overtaking elastic
pairwise interactions between solitons. In this paper, we extend this
theory to soliton gases in bidirectional integrable Eulerian systems
where both head-on and overtaking collisions of solitons take place.
We distinguish between two qualitatively different types of
bidirectional soliton gases: isotropic gases, in which the position
shifts accompanying the head-on and overtaking soliton collisions have
the same sign, and anisotropic gases, in which the position shifts for
head-on and overtaking collisions have opposite signs.  We construct
kinetic equations for both types of bidirectional soliton gases and
solve the respective shock-tube problems for the collision of two
``monochromatic'' soliton beams consisting of solitons of
approximately the same amplitude and velocity. The corresponding weak
solutions of the kinetic equations consisting of differing uniform
states separated by contact discontinuities for the mean flow are
constructed. Concrete examples of bidirectional Eulerian soliton gases
for the defocusing nonlinear Schr\"odinger (NLS) equation and the
resonant NLS equation are considered.  The kinetic equation of the
resonant NLS soliton gas is shown to be equivalent to that of the
shallow-water bidirectional soliton gas described by the
Kaup-Boussinesq equations. The analytical results for shock-tube
Riemann problems for bidirectional soliton gases are shown to be in
excellent agreement with direct numerical simulations.
\end{abstract}

\maketitle

\section{Introduction} 
\label{sec:intro}

Dispersive hydrodynamics modeled by hyperbolic conservation laws
regularized by conservative, dispersive corrections describe various
nonlinear wave structures that include solitary waves (solitons),
dispersive shock waves (DSWs), rarefaction waves and their
interactions~\cite{biondini_dispersive_2016}.  A particular feature of
dispersive hydrodynamics is the intrinsic scale separation, often
providing a qualitatively new perspective on some classical
mathematical and fluid dynamical settings (such as Riemann problems or
flows past topography), but also revealing novel phenomena such as
hydrodynamic soliton tunneling~\cite{maiden_solitonic_2018,
sprenger_hydrodynamic_2018} and expansion
shocks~\cite{el_expansion_2016}.

On a small-scale, microscopic, level dispersive hydrodynamics
typically involve coherent nonlinear wave structures such as solitons
and rapidly oscillating periodic waves, while the large-scale,
macroscopic coherent features are represented by slow modulations of
these periodic waves or soliton trains. The prominent example of a
dispersive hydrodynamic structure exhibiting such two-scale coherence
and persisting in integrable and non-integrable systems is DSW, the
dispersive analog of a classical, viscous shock
wave~\cite{el_dispersive_2016}.

There is another class of problems in dispersive hydrodynamics, which
involve the wave structures exhibiting coherence at a microscopic
scale, while being macroscopically incoherent, in the sense that the
values of the wave field at two points separated by a distance much
larger than the intrinsic dispersive length of the system (the soliton
width), are not dynamically related.  These structures can be broadly
viewed as dispersive-hydrodynamic analogs of turbulence, and the
qualitative and quantitative properties of such a conservative
turbulence strongly depend on the integrability properties of the
underlying microscopic dynamics.  In~\cite{zakharov_turbulence_2009}
V.E.~Zakharov introduced the notion of ``integrable turbulence'' for
random nonlinear wave fields governed by integrable equations such as
the Korteweg-de Vries (KdV) or Nonlinear Schr\"odinger (NLS)
equations. The source of randomness in integrable turbulence is
typically related to some sort of stochastic initial or boundary
conditions although one can envisage dynamical mechanisms of the
effective randomization of the wave
field~\cite{gurevich_development_1999, el_dam_2016}.  The theoretical
perspective of integrable turbulence has turned out to be very
fruitful, providing new insights into some long-standing problems of
nonlinear physics related e.g. to modulational instability and the
formation of rogue waves~\cite{Walczak:15,Randoux:16b,
gelash_strongly_2018}. Indeed, integrable turbulence proved a
promising theoretical framework for the interpretation of experimental
and observational data in fiber optics and fluid
dynamics~\cite{costa_soliton_2014}.

Solitons, viewed as stable ``wave-particles'' of macroscopic
dispersive-hydrodynamic structures, can form large disordered,
statistical ensembles, strikingly different from the macroscopically
coherent DSWs, and calling for the analogy with gases of classical or
quantum particles.  Such statistical soliton ensembles, or ``soliton
gases'', can be naturally generated from both non-vanishing
deterministic (e.g. periodic or quasiperiodic) and random initial
conditions due to the processes of soliton
fissioning~\cite{trillo_experimental_2016,trillo_observation_2016} or
modulation instability~\cite{gelash_bound_2019}. The ubiquity of
solitons in applications and the integrable nature of the underlying
wave dynamics makes soliton gases a particularly attractive object for
modeling the complex nonlinear wave phenomena occurring in the ocean
and in high-intensity incoherent light propagation through optical
materials, see~\cite{el_spectral_2020} and references therein. The
random nonlinear wave field in a soliton gas represents a particular
case of integrable turbulence~\cite{zakharov_turbulence_2009}.

Within the inverse scattering transform (IST) formalism each soliton
is characterized by a discrete eigenvalue $\lambda_j$ of the spectrum
of the linear operator associated with the integrable nonlinear
evolution equation. There are two basic aspects of the microscopic,
soliton dynamics that determine the macroscopic, statistical
properties of integrable soliton gases/turbulence: (i) isospectrality
of integrable evolution resulting in the preservation of soliton
eigenvalues; (ii) pairwise elastic collisions accompanied by
phase/position shifts expressed in terms of the respective spectral
parameters of the interacting solitons.

The macroscopic properties of a soliton gas are determined by the
spectral characteristics called the density of states (DOS)
$f(\lambda)> 0$, defined such that the number of solitons found at the
moment of time $t$ in the element
$[\lambda, \lambda+d\lambda] \times [x, x+dx] $ of the phase space is
$f(\lambda)d\lambda d x$ (assuming $\lambda \in \mathbb{R}$, the
generalization to complex spectrum being
straightforward~\cite{el_kinetic_2005}).  DOS represents the
definitive statistical characteristics of soliton gas distinguishing
it from an arbitrary random collection of solitons.  The first
controlled generation of soliton gas characterized by a measurable DOS
has been recently reported in~\cite{suret_nonlinear_2020}.

For uniform, statistically homogeneous soliton gases the DOS depends
on the spectral parameter only. For spatially nonhomogeneous soliton
gases one has $f \equiv f(\lambda, x, t)$, and the isospectrality of
integrable evolution implies the conservation equation
\begin{equation}\label{eq:conserv}
f_t+(sf)_x=0,
\end{equation}
where the transport velocity (the mean velocity of a ``tracer''
soliton in a gas) $s(\lambda, x,t)$ is found from the integral
equation of state~\cite{el_kinetic_2005}
\begin{align}\label{eq:state}
s(\lambda, x,t)= c(\lambda) + \int_{\Omega}
&\Delta(\lambda, \mu) f(\mu, x,t) \notag\\
&\times |s(\lambda,x,t) - s(\mu, x,t)|d\mu.
\end{align}
Here $c(\lambda)$ is the velocity of an isolated single soliton with the
spectral parameter $\lambda \in \Omega$, and the integral term
describes its modification due to collisions with other,
``$\mu$-solitons'' in a gas, each collision being accompanied by the
position shift $\Delta(\lambda, \mu)$, often called the phase shift.
The integration in~\eqref{eq:state} is performed over the spectral
support $ \Omega \subset \mathbb{R}$ of the DOS $f(\lambda, x, t)$.
If one assumes that (i)
$\sgn[\Delta(\lambda, \mu)] = \pm \sgn(\lambda - \mu)$; and (ii)
$s'(\lambda)\ne 0$ the modulus sign in~\eqref{eq:state} can be removed
by introducing
$\Delta(\lambda, \mu) = \sgn(\lambda - \mu) G(\lambda, \mu)$ so that
one arrives at the conventional form of the equation of state as
in~\cite{el_kinetic_2005, el_kinetic_2011}, involving
$G(\lambda, \mu)$ rather than $\Delta(\lambda, \mu)$ as the integral
kernel. E.g. for the KdV solitons one has
$\sgn[\Delta(\lambda, \mu)]=+\sgn(\lambda - \mu)$, $s'(\lambda)>0$ and
$G(\lambda, \mu)=\lambda^{-1}\ln |(\lambda - \mu)/(\lambda+ \mu)| $
(see e.g.~\cite{ablowitz_note_1982}).

The transport equation~\eqref{eq:conserv} for the DOS complemented by
the integral equation of state~\eqref{eq:state} comprise the kinetic
equation for soliton gas. Kinetic equation of the
type~\eqref{eq:conserv},\eqref{eq:state} was first introduced
in~\cite{zakharov_kinetic_1971} for the case of rarefied, or dilute,
gas of KdV solitons, when the interaction term in the equation
state~\eqref{eq:state} represents a small correction and the soliton
velocity in a gas is found from the expression
$s \approx 4\lambda^2 + \lambda^{-1} \int_0^{\lambda_{\max}} \ln
|(\lambda - \mu)/(\lambda+ \mu)|f(\lambda,x,t)[4\lambda^2 - 4 \mu^2] d
\mu$, which is an approximate solution of the equation of
state~\eqref{eq:state} for the KdV soliton gas.  The full kinetic
equation~\eqref{eq:conserv},\eqref{eq:state} for a dense soliton gas
was derived and analyzed in the context of the KdV equation
in~\cite{el_thermodynamic_2003,el_critical_2016} and the focusing NLS
equation in~\cite{el_kinetic_2005, el_spectral_2020} (in the latter
case $\lambda \in \mathbb{C}$).  A general mathematical analysis of
the kinetic equation~\eqref{eq:conserv},\eqref{eq:state} has been
undertaken in~\cite{el_kinetic_2011}, which showed that it possesses
an infinite series of integrable linearly degenerate hyperbolic
reductions.  Very recently the kinetic
equation~\eqref{eq:conserv},\eqref{eq:state} has attracted much
attention in the context of generalized hydrodynamics, a statistical
theory of quantum many-body integrable systems,
see~\cite{doyon_soliton_2018, doyon_geometric_2018, vu_equations_2019}
and references therein.

In the context of dispersive hydrodynamics the kinetic
equation~\eqref{eq:conserv},\eqref{eq:state} describes
``unidirectional'' soliton gases supported by scalar integrable
equations of the form
\begin{equation}
\label{eq:scalar_dh}
u_t + F(u)_x =  (D[u])_x , 
\end{equation}
where $F(u)$ is the nonlinear hyperbolic flux and $D[u]$ is a
differential (generally integro-differential) operator, possibly
nonlinear, that gives rise to a real-valued linear dispersion
relation. The spectral single-soliton solutions to
Eq.~\eqref{eq:scalar_dh} are characterized by the soliton velocity
$c(\lambda)$ and the phase shift kernel $\Delta(\lambda, \mu)$
characterizing the ``overtaking'' two-soliton interactions.  However,
the scalar integrable dispersive hydrodynamics of the
form~\eqref{eq:scalar_dh}, such as the KdV, modified KdV, Camassa-Holm
or Benjamin-Ono equations typically arise as small-amplitude,
``unidirectional'' approximations of more general Eulerian
bidirectional systems (see~\cite{whitham_linear_1999})
\begin{equation}
\label{eq:disp_Euler}
\begin{split}
\rho_t + (\rho u)_x &= (D_1[\rho,u])_x\, , \\
(\rho u)_t + \left ( \rho u^2 + P(\rho) \right )_x &=
(D_2[\rho,u])_x \,,
\end{split}
\end{equation}
where $D_{1,2}[\rho,u]$ are conservative, dispersive operators,
$P(\rho)>0$ is the monotonically increasing pressure law, and $\rho$,
$u$ are interpreted as a mass density and fluid velocity,
respectively.  This class of equations generalizes the shallow water
and isentropic gas dynamics equations while encompassing many of the
integrable dispersive hydrodynamic models such as the Kaup-Boussinesq
(KB) system~\cite{kaup_higher_1975}, the hydrodynamic form of the
defocusing NLS equation~\cite{kamchatnov_nonlinear_2000} or the
Calogero-Sutherland system describing the dispersive hydrodynamics of
quantum many-body systems~\cite{abanov_integrable_2009}.  Due to the
bidirectional nature, the Eulerian dispersive
hydrodynamics~\eqref{eq:disp_Euler} supports solitons that experience
both overtaking and head-on elastic collisions which are generally
characterized by two different phase shift kernels
$\Delta_1(\lambda, \mu) \ne \Delta_2(\lambda, \mu)$.  Indeed, the
rarefied bidirectional shallow-water soliton gas realized in the water
tank experiments~\cite{redor_experimental_2019,redor_analysis_2020}
was modeled by the KB system~\cite{kaup_higher_1975},
which exhibits qualitatively different properties for head-on and
overtaking position shifts in the pairwise soliton
collisions~\cite{li_bidirectional_2003} so that the overtaking
interactions can be characterized as ``strong'' and the head-on
interactions as ``weak''. We shall term such collisions and the
associated soliton gases as ``anisotropic''. On the other hand, some
bidirectional dispersive hydrodynamic systems support soliton
solutions that exhibit ``isotropic'' collisions characterized by the
same phase shift kernel $\Delta(\eta, \mu)$ for the head-on and
overtaking interactions (e.g. the defocusing NLS
equation~\cite{zakharov_interaction_1973}).

Despite the significant recent advances of the kinetic theory of
unidirectional soliton gases, a consistent general extension of this
theory to the physically important bidirectional case has not been
available so far, and this paper is devoted to the development of such
an extension. The paper is organized as follows. In
Section~\ref{sec:bidir} we present the general construction of the
kinetic equation for bidirectional soliton gas and realize it for the
cases of the defocusing nonlinear Schr\"odinger (DNLS) equation and
its ``stable'' negative dispersion counterpart, the so-called resonant
NLS (RNLS) equation, having applications in magneto-hydrodynamics of
cold collisionless plasma~\cite{lee_resonant_2007}, and reducible to
the KB system for shallow-water waves by a simple change
of variables.  It turns out that, due to the pairwise collisions of
dark DNLS solitons being isotropic, the bidirectional kinetic equation
for the dark (grey) solitons of the DNLS equation reduces to the
unidirectional kinetic equation of the
form~\eqref{eq:conserv},\eqref{eq:state}. Contrastingly, the soliton
collisions of anti-dark RNLS solitons are anisotropic, and the kinetic
equation for this case represents a pair of the kinetic equations of
the type~\eqref{eq:conserv},\eqref{eq:state} with some nonlinear
coupling through the equation of state.  In Section~\ref{sec:mean} we
derive expressions for the mean field in both soliton gases in terms
of the spectral DOS. To demonstrate the efficacy of the developed
theory we consider in Section~\ref{sec:beam} the ``shock-tube''
Riemann problem describing the collision of ``monochromatic'' soliton
beams for both types of bidirectional gases.  The collisions are
described by weak solutions to the bidirectional kinetic equations,
consisting of a number differing constant states for the DOS,
separated by contact discontinuities for the component densities,
satisfying appropriate Rankine-Hugoniot conditions. The analytical
results are shown to be in an excellent agreement with direct
numerical simulations of the soliton gas shock-tube problem for DNLS
and RNLS equations.

\section{Kinetic equation for bidirectional soliton gas}
\label{sec:bidir}

In this section we derive the kinetic equation for integrable Eulerian
dispersive hydrodynamics~\eqref{eq:disp_Euler} using the general
physical construction proposed in~\cite{el_kinetic_2005} for a
uni-directional case. The construction uses an extension of the
original Zakharov's phase shift
reasoning~\cite{zakharov_kinetic_1971}, which, strictly speaking, is
applicable only in a rarefied gas case. However, the resulting kinetic
equation~\eqref{eq:conserv},\eqref{eq:state} turns out to provide
correct description for a dense gas, mathematically justified by the
thermodynamic limit of the finite-gap Whitham modulation systems for
the cases of the KdV~\cite{el_thermodynamic_2003} and the focusing
NLS~\cite{el_spectral_2020} equations.  Our results for bidirectional
gas will be later supported by comparisons with direct numerical
simulations of the relevant soliton gases, justifying the validity of
the phenomenological derivation.

\subsection{Isotropic and anisotropic bidirectional soliton gases}

Suppose that the system~\eqref{eq:disp_Euler} supports a family of
bidirectional soliton solutions that bifurcate from the two branches
of the linear wave spectrum $\omega=\omega_{\pm}(k)$
of~\eqref{eq:disp_Euler} so that $\omega_-(k)/k < \omega_+(k)/k$ in
the long wavelength limit $k \to 0$.  We denote the corresponding
soliton families $(\rho^-_{\rm s},u^-_{\rm s})$ and
$(\rho^+_{\rm s},u^+_{\rm s})$.  Let these soliton solutions be
parameterized by a real-valued spectral (IST) parameter $\lambda$ so
that $\lambda \in \Omega_+ $ for the ``fast'' branch and
$\lambda \in \Omega_-$ for the ``slow'' branch, where $\Omega_\pm$ are
simply-connected subsets of $\mathbb{R}$ with one intersection point
at most.  Let the respective soliton velocities be
$c_\pm(\lambda)$. For convenience we assume that
$c_{\pm}'(\lambda)>0$, and $c_-(\lambda_1) \ < c_+(\lambda_2)$ if
$\lambda_1 \in \Omega_-$ and $\lambda_2 \in \Omega_+$,
$\lambda_1 \ne \lambda_2$.  If $\Omega_- \cap \Omega_+= \{\lambda_*\}$
we assume $c_-(\lambda_*)=c_+(\lambda_*)$. The above assumptions are
consistent with all concrete examples of integrable dispersive
hydrodynamics we consider in this paper.

One can distinguish between two types of the pairwise collisions in a
bidirectional soliton gas: the overtaking collisions between
solitons belonging to the same spectral branch and characterized by
the position shifts $\Delta_{++}$ and $\Delta_{--}$ respectively, and
the ``head-on'' collisions between solitons of different branches,
characterized by the position shifts $\Delta_{+ -}$ and $\Delta_{-+}$.
Let $\lambda \ne \mu$, and $\Delta_{\pm \pm} (\lambda,\mu)$ and
$\Delta_{\pm \mp} (\lambda,\mu)$ denote the position shifts of a
$\lambda$-soliton due to its collision with a $\mu$-soliton, with the
first and the second signs $\pm$ in the subscript indicating the
branch belonging of the $\lambda$-soliton and the $\mu$-soliton
respectively, e.g. $\Delta_{-+}(\lambda, \mu)$ is the position shift
of a $\lambda$-soliton with $\lambda \in \Omega_-$ in a collision with
a $\mu$-soliton with $\mu \in \Omega_+$.

We call the bidirectional soliton gas ``isotropic'' if the position
shifts for the overtaking and head-on collisions between $\lambda$-
and $\mu$- solitons satisfy the following sign conditions:
\begin{equation}\label{sign_cond}
\sgn [\Delta_{++}]= \sgn[\Delta_{+-}], \quad  \sgn[\Delta_{--}]=
\sgn[\Delta_{-+}],
\end{equation}
i.e.  the $\lambda$-soliton experiences a shift of a certain sign, say
shift forward (and the $\mu$-soliton--- the shift of an opposite sign)
irrespectively of the type of the collision---overtaking or head-on.
If conditions~\eqref{sign_cond} are not satisfied, i.e. the sign of
the phase shift depends on the type of the collision, we shall call
the corresponding soliton gas ``anisotropic''.  The difference
between isotropic and anisotropic collisions is illustrated in
Fig.~\ref{fig:phase_shift} using concrete examples.

\subsection{Kinetic equation for bidirectional soliton gas:
general construction}
\label{sec:general}

Following the construction for unidirectional soliton gas outlined in
the introduction, we now consider bidirectional soliton gases for
integrable Eulerian equations~\eqref{eq:disp_Euler}. We introduce two
separate DOS's $f_-(\lambda,x,t)$ and $f_+(\lambda,x,t)$ for the
populations of solitons whose spectral parameters belong to the
slow ($\Omega_-$) and fast ($\Omega_+$) branches of the
spectral set $\Omega$ respectively.  The isospectrality of integrable
evolution implies now two separate conservation laws:
\begin{equation}
\label{eq:kin2}
(f_-)_t + (s_- f_-)_x = 0,\quad (f_+)_t + (s_+ f_+)_x = 0,
\end{equation}
where $s_-(\lambda,x,t)$ and $s_+(\lambda,x,t)$ are the transport
velocities associated with the motion of slow solitons and fast
solitons associated with $\Omega_-$ and $\Omega_+$ branches
respectively.  We derive the equations of state for $s_\pm$ using the
direct phenomenological approach proposed~\cite{el_kinetic_2005}: we
identify $s_\pm(\lambda,x,t)$ as the velocity of a trial
$\lambda$-soliton of the gas. Consider, for instance, a tracer
$\lambda$-soliton from the slow branch, $\lambda \in \Omega_-$, and
compute its displacement in a gas over the ``mesoscopic'' time
interval $dt$, sufficiently large to incorporate a large number of
collisions, but sufficiently small to ensure that the spatiotemporal
field $f_\pm(\lambda,x,t)$ is stationary over $dt$ and homogeneous on
a typical spatial scale $c_\pm(\lambda) dt$. Having this in mind, we
drop the $(x,t)$-dependence for convenience.  Each overtaking
collision with a soliton of the same branch $\mu \in \Omega_-$ shifts
the $\lambda$-soliton by the distance $\Delta_{--}(\lambda,\mu)$.
Thus the displacement of the $\lambda$-soliton over the time $dt$ due
to the overtaking collisions is given by
$\int_{\Omega_-} \Delta_{--}(\lambda,\mu) f_-(\mu)
|s_-(\lambda)-s_-(\mu)| dt\, d\mu$ where
$f_-(\mu) |s_-(\lambda)-s_-(\mu)| dt$ is the average number of
collisions with encountered $\mu$-solitons
(cf.~\cite{el_kinetic_2005}). Additionally, each head-on collision
with a fast soliton $\mu \in \Omega_+$ shifts the slow
$\lambda$-soliton with $\lambda \in \Omega_-$ by
$\Delta_{-+}(\lambda,\mu)$, and the resulting displacement after a
time $dt$ is
$\int_{\Omega_+}\Delta_{-+}(\lambda,\mu) f_+(\mu)
|s_+(\lambda)-s_-(\mu)| dt\, d\mu$. A similar consideration is applied
to the fast soliton branch, $\lambda \in \Omega_+$, in the gas.
Equating the total displacements of the slow and fast
$\lambda$-solitons to $s_-(\lambda) dt$ and $s_+(\lambda) dt$
respectively, we obtain the equation of state of a bidirectional gas
in the form of two coupled linear integral equations:
\begin{widetext}
\begin{equation}
\label{eq:s2}
\begin{split}
&s_-(\lambda) = c_-(\lambda) + \int_{\Omega_-} 
\Delta_{--} (\lambda,\mu) f_-(\mu) |s_-(\lambda)-s_-(\mu)| d \mu+
\int_{\Omega_+}  \Delta_{-+} (\lambda,\mu) f_+(\mu)
|s_-(\lambda)-s_+(\mu)| d\mu, \\
&s_+(\lambda) = c_+(\lambda) + \int_{\Omega_+} 
\Delta_{++} (\lambda,\mu) f_+(\mu) |s_+(\lambda)-s_+(\mu)| d \mu +
\int_{\Omega_-}  \Delta_{+-} (\lambda,\mu) f_-(\mu)
|s_+(\lambda)-s_-(\mu)| d\mu,
\end{split}
\end{equation}
\end{widetext}
where $\lambda \in \Omega_-$ for the first equation and
$\lambda \in \Omega_+$ for the second equation.  If the spectral
support $\Omega = \Omega_- \cup \Omega_+ \subset \mathbb{R}$ is a
simply connected set and the gas is isotropic, the distinction between
the fast and slow branches becomes unnecessary and the kinetic
equation~\eqref{eq:kin2},\eqref{eq:s2} for bidirectional soliton gas
is naturally reduced to the unidirectional gas
equation~\eqref{eq:conserv},\eqref{eq:state} for a single DOS
$f(\lambda)$ defined on the entire set $\Omega$.  We will show in
Sec.~\ref{sec:beam}, using concrete examples, that the dynamics
governed by the kinetic equations~\eqref{eq:conserv},\eqref{eq:state}
and~\eqref{eq:kin2},\eqref{eq:s2} is in a very good agreement with the
results of direct numerical simulations of isotropic and anisotropic
bidirectional soliton gases respectively.

\subsection{Kinetic equation for bidirectional soliton gas:
examples}
\label{sec:ex}

As a representative (and physically relevant) example, we consider the
integrable Eulerian dispersive hydrodynamics
\begin{equation}
\label{eq:NLS}
\begin{split}
\rho_t+(\rho u)_x &=0,\\
(\rho u)_t+\left(\rho u^2 + \frac{\rho^2}{2} \right)_x &=
\frac{\sigma}{4} \left[ \rho \left ( \ln{\rho} \right )_{xx}
\right]_x \, , \quad \sigma = \pm 1 \,.
\end{split}
\end{equation}
For $\sigma=1$, system~\eqref{eq:NLS} is equivalent to the defocusing
nonlinear Schr\"odinger (DNLS) equation:
\begin{equation}
\label{eq:DNLS}
i \psi_t + \tfrac12 \psi_{xx}-|\psi|^2 \psi=0, \quad \psi=\sqrt{\rho}
\exp \left(i \int u \,dx \right).
\end{equation}
The DNLS equation has a number of physical applications. In particular
it describes propagation of light beams through optical fibers in the
regime of normal dispersion, as well as nonlinear matter waves in
quasi-1D repulsive Bose-Einstein condensates (BECs), see for
instance~\cite{yang_nonlinear_2010}. Pertinent to the
present context, rarefied gas of dark solitons in quasi-1D BEC has
been investigated in \cite{schmidt_non-thermal_2012,
wang_transitions_2015}.

The DNLS equation has a family of dark (or
grey) soliton solutions~\cite{zakharov_interaction_1973}
\begin{equation}
\label{eq:dark}
\begin{split}
&\rho_{\rm s}^\pm = 1- (1-\lambda^2){\rm sech}^2
\big[\sqrt{1-\lambda^2}(x- c_\pm t) \big],\\
&u_{\rm s}^\pm =
\lambda\left(1-\frac{1}{\rho_{\rm s}^\pm(x,t)} \right),\quad c_\pm =
\lambda \in \Omega_\pm,
\end{split}
\end{equation}
where $\Omega_- = (-1,0]$ for the slow solitons branch and
$\Omega_+=[0,+1)$ for the fast solitons branch; note that solutions
$(\rho_{\rm s}^+,u_{\rm s}^+)$ and $(\rho_{\rm s}^-,u_{\rm s}^-)$ have
the same analytical expression.

\begin{figure}[h]
\centering
\includegraphics{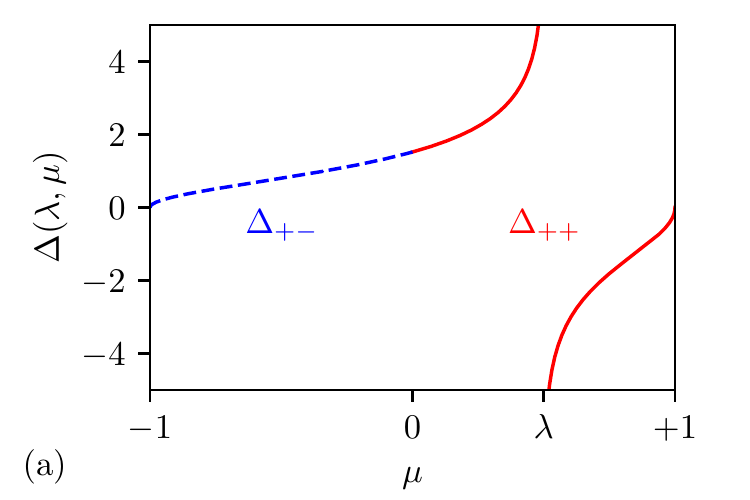}
\includegraphics{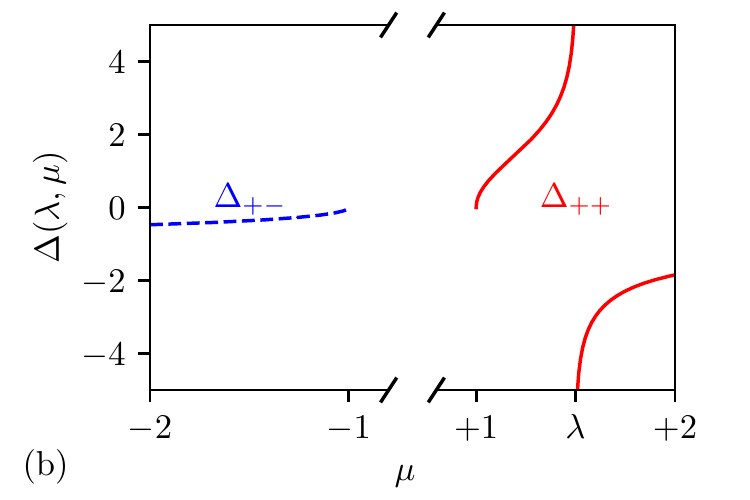}
\caption{Variation of the phase shifts in the isotropic~(a) and
anisotropic~(b) interactions of solitons with spectral parameters
$\lambda$ and $\mu$. The $\lambda$-soliton belongs to the ``+''-branch
with: $\lambda=1/2$~(a) and $\lambda=3/2$~(b). Solid lines represent
the variation of $\Delta_{++}$ and dashed lines the variation of
$\Delta_{+-}$.}
\label{fig:phase_shift}
\end{figure}

\begin{figure}[h]
\centering
\includegraphics{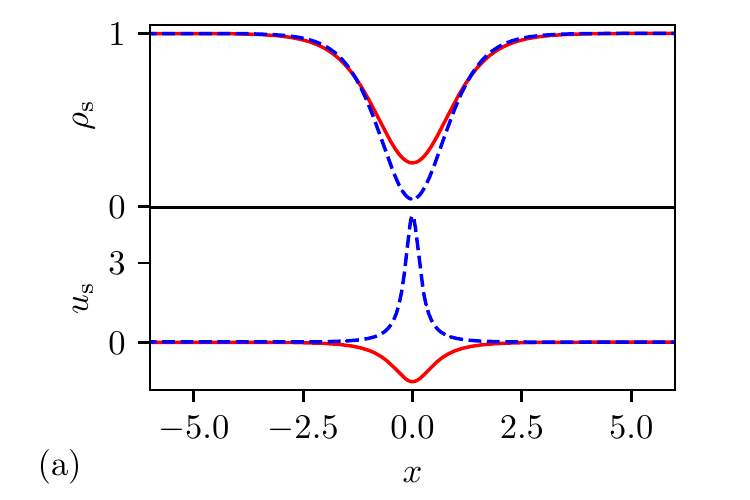}
\includegraphics{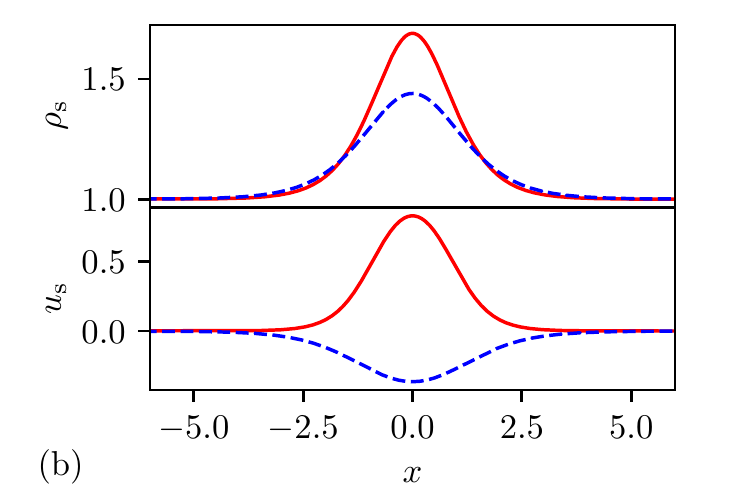}
\includegraphics{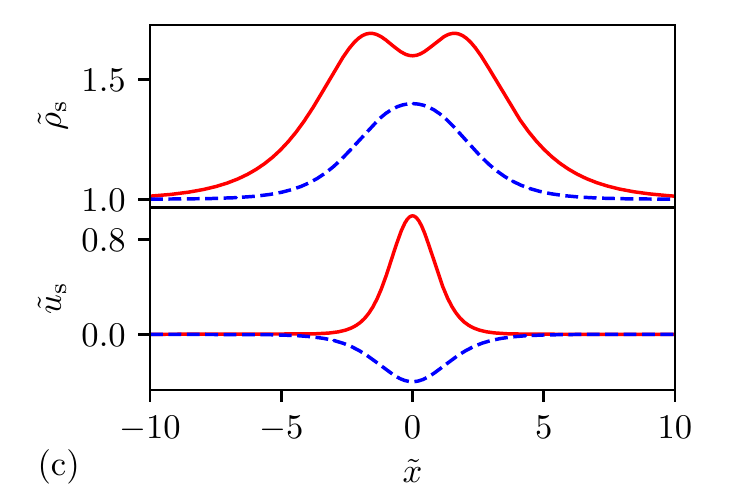}
\caption{Soliton solutions: solid lines represent fast branch
solutions~$(\rho_{\rm s}^+, u_{\rm s}^+)$ and dashed lines slow branch
solutions~$(\rho_{\rm s}^-, u_{\rm s}^-)$. (a)~Dark soliton solutions
of the DNLS equation~\eqref{eq:dark} with $\lambda =
+0.5,-0.2$. (b)~Anti-dark soliton solutions of the RNLS
equation~\eqref{eq:sol_RNLS} with $\lambda = +1.3,-1.2$. (c)~Anti-dark
soliton solutions of the KB system~\eqref{eq:sol_KB} with
$\lambda = +1.3,-1.2$.}
\label{fig:soliton}
\end{figure}

Without loss of generality we assumed in~\eqref{eq:dark} the unit
density background. Typical dark soliton solutions are displayed in
Fig.~\ref{fig:soliton}.  The position shifts in the DNLS overtaking
and head-on soliton collisions are given by the same analytical
expression
$\Delta_{\pm\pm} (\lambda, \mu)= \Delta_{\pm\mp} (\lambda, \mu) \equiv
\Delta(\lambda, \mu)$, where
\begin{equation}
\label{eq:Delta_DNLS}
\begin{split}
\Delta(\lambda,\mu) &= \sgn(\lambda-\mu)  \, G_1(\lambda, \mu), \\
 G_1(\lambda, \mu) &\equiv \tfrac{1}{2 \sqrt{1-\lambda^2}}
\log \tfrac{(\lambda-\mu)^2 + \big(\sqrt{1-\lambda^2}+\sqrt{1-\mu^2}
\big)^2} {(\lambda-\mu)^2 + \big(\sqrt{1-\lambda^2}-\sqrt{1-\mu^2}
\big)^2},
\end{split}
\end{equation}
for all $\lambda, \mu \in (-1, 1)$.  One can verify that the soliton
position shifts given by~\eqref{eq:Delta_DNLS} satisfy the isotropy
conditions~\eqref{sign_cond}. The variation of $\Delta(\lambda,\mu)$
with respect to $\mu$ for a fixed $\lambda$ is displayed in
Fig.~\ref{fig:phase_shift}. One can see that the position shifts for
the head-on and overtaking collisions lie on the same curve with
$\Delta(\lambda, \mu)$ being continuous at $\lambda=0$, the point of
intersection of $\Omega_-$ and $\Omega_+$.  Due to the isotropic
nature of the DNLS soliton interactions the coupled kinetic
equation~\eqref{eq:kin2},\eqref{eq:s2} for the bidirectional DNLS gas
reduces to the single kinetic equation~\eqref{eq:conserv} with the
equation of state
\begin{align}
\label{eq:s_DNLS}
s(\lambda,x,t)=\lambda + \int \limits_{-1}^{+1} &G_1(\lambda,
\mu)f(\mu,x,t) \notag\\
&\times (s(\lambda,x,t)- s(\mu,x,t)) d \mu,
\end{align}
where $\lambda \in (-1,1)$ and with the assumption that
$s'(\lambda)>0$; the latter assumption is verified by comparison to
numerics in Sec.~\ref{sec:riem}.  This reduction to the unidirectional
case is similar to the kinetic equation derived
in~\cite{el_spectral_2020} for the bidirectional soliton and breather
gases of the focusing NLS equation which also exhibits isotropic
soliton/breather collisions, with the essential difference that the
integration in the focusing NLS case occurs over a compact domain in a
complex plane of the spectral parameter.

For $\sigma=-1$ the system~\eqref{eq:NLS} is equivalent to the
so-called resonant NLS (RNLS) equation (see,
e.g.,~\cite{lee_solitons_2007})
\begin{equation}
\label{eq:RNLS}
\begin{split}
&i \psi_t + \tfrac12 \psi_{xx}-|\psi|^2 \psi=|\psi|_{xx} \psi/|\psi|,\\
&\psi=\sqrt{\rho}\exp \left(i \int u \,dx \right).
\end{split}
\end{equation}
This equation, in particular,
describes long magneto-acoustic waves in a cold plasma propagating
across the magnetic field~\cite{gurevich_origin_1988}.
The RNLS equation has a family of anti-dark
soliton solutions given by~\cite{lee_solitons_2007} 
\begin{equation}
\label{eq:sol_RNLS}
\begin{split}
&\rho_{\rm s}^\pm = 1+ (\lambda^2-1){\rm sech}^2
\big[\sqrt{\lambda^2-1}(x-c_\pm
t) \big],\\
&u_{\rm s}^\pm = \lambda\left(1-\frac{1}{\rho_{\rm s}^\pm(x,t)}
\right),\quad c_\pm = \lambda \in \Omega_\pm.
\end{split}
\end{equation}
Solutions $(\rho_{\rm s}^+,u_{\rm s}^+)$ and $(\rho_{\rm s}^-,u_{\rm
s}^-)$ have the same analytical expression. Typical anti-dark soliton
solutions are displayed in Fig.~\ref{fig:soliton}.

In contrast with the DNLS system, the spectral set $\Omega$ of the
RNLS soliton is spanned by two disconnected subsets: $\Omega_- =
(-\infty,-1)$ for slow solitons and $\Omega_+ = (+1,+\infty)$. Similar
to the DNLS equation, the position shifts in head-on and overtaking
collisions are given by the same analytical expression $ \Delta_{\pm
\pm}(\lambda, \mu)= \Delta_{\pm \mp}(\lambda, \mu) \equiv
\Delta(\lambda, \mu)$, where
\begin{equation}
\label{eq:DeltaRNLS}
\begin{split}
\Delta(\lambda, \mu)&= \sgn(\lambda-\mu) \,
G_2(\lambda, \mu), \\
G_2(\lambda, \mu) &\equiv \tfrac{1}{2
\sqrt{\lambda^2-1}}  \log \tfrac{(\lambda-\mu)^2 -
\big(\sqrt{\lambda^2-1}+\sqrt{\mu^2-1} \big)^2} {(\lambda-\mu)^2 -
\big(\sqrt{\lambda^2-1}-\sqrt{\mu^2-1} \big)^2}.
\end{split}
\end{equation}
However, one can verify that, unlike in the DNLS case, the
isotropy condition~\eqref{sign_cond} is not satisfied. Indeed, it
follows from~\eqref{eq:DeltaRNLS} that
$\sgn[\Delta_{\pm \pm} (\lambda,\mu)] = \sgn(\lambda -\mu)$, whereas
$\sgn[\Delta_{\pm \mp} (\lambda,\mu)] = -\sgn(\lambda -\mu)$, that is
in a head-on collision between a $\lambda$-soliton and a $\mu$-soliton
with $\lambda>\mu$, the $\lambda$-soliton's position is now shifted
backward.  The variation of $\Delta_{\pm \pm}(\lambda, \mu)$ for the
RNLS equation shown in Fig.~\ref{fig:phase_shift}. One can see that
it is qualitatively different from the variation of
$\Delta_{\pm \mp}(\lambda, \mu)$ for the DNLS equation.

The kinetic equation for the anisotropic RNLS soliton gas has then the
form of two continuity equations~\eqref{eq:kin2} complemented by the
coupled equations of state
\begin{widetext}
\begin{equation}
\label{eq:s_RNLS}
\begin{split}
&s_-(\lambda) = \lambda + \int \limits_{-\infty}^{-1} G_2(\lambda,
\mu) f_-(\mu) (s_-(\lambda)-s_-(\mu)) d \mu + \int
\limits_{+1}^{\infty} G_2(\lambda, \mu) f_+(\mu)
(s_-(\lambda)-s_+(\mu)) d\mu,\\
&s_+(\lambda) = \lambda + \int \limits_{+1}^{+\infty} G_2(\lambda,
\mu) f_+(\mu) (s_+(\lambda)-s_+(\mu)) d \mu + \int
\limits_{-\infty}^{-1} G_2(\lambda, \mu) f_-(\mu)
(s_+(\lambda)-s_-(\mu)) d\mu,
\end{split}
\end{equation}
\end{widetext}
with the assumptions that $s_\pm'(\lambda)>0$ and $s_+>s_-$; the
latter assumption is verified by direct comparison with numerics in
Sec.~\ref{sec:riem}.

We note that the change of variables:
\begin{equation}
\label{eq:changeKB}
\begin{split}
&\tilde \rho = \rho + \frac12 \left( u + \frac{\rho_x}{2\rho}
\right)_x,\quad
\tilde u = u + \frac{\rho_x}{2\rho},\\
&\tilde x = \frac{2}{\sqrt 3} \,x,\quad \tilde t =  \frac{2}{\sqrt 3}
\,t,
\end{split}
\end{equation}
transforms the RNLS equation into the KB
system~\cite{kaup_higher_1975}:
\begin{equation}
\label{eq:KB}
\begin{split}
\tilde\rho_{\tilde t}+ (\tilde \rho \tilde u )_{\tilde x} =- \frac13
\tilde u_{\tilde x \tilde x \tilde x},\quad
\tilde u_{\tilde t}+\tilde u \tilde u_{\tilde x} + \tilde \rho_{\tilde
x} = 0,
\end{split}
\end{equation}
describing bidirectional shallow water waves.
The soliton solution of~\eqref{eq:KB} is obtained by
transforming~\eqref{eq:sol_RNLS} with the change of
variables~\eqref{eq:changeKB} (cf. Appendix~\ref{sec:sol_KB}), and the
phase of a $\lambda$-soliton after colliding with a $\mu$-soliton is:
$2/\sqrt 3 \times \sgn(\lambda-\mu) G_2(\lambda,\mu)$. Thus the the
RNLS and the KB soliton gas share the same anisotropic kinetic
description. One can notice in Fig.~\ref{fig:soliton} that the bimodal
soliton of KB system transforms into a unimodal soliton of the RNLS
equation with the change of variables~\eqref{eq:changeKB}.
In the numerical examples presented in the next section we will mostly
focus on the anisotropic RNLS soliton gas for a direct comparison with
the isotropic DNLS soliton gas.

\section{Ensemble averages of the wave field in bidirectional soliton gases}
\label{sec:mean}

The DOS $f(\lambda)$ ($f_\pm(\lambda)$ in the anisotropic case)
represents a comprehensive spectral characteristics, that, in
principle, determines all statistical parameters of the nonlinear
random wave field $(\rho(x,t),u(x,t))$ in a soliton gas.  The most
obvious set of such statistical parameters are the ensemble averages
of the conserved quantities. We note that for the KdV soliton gas the
averages $\langle u \rangle$, $\langle u^2 \rangle$ were determined in
terms of DOS in~\cite{el_soliton_2001, el_critical_2016} using the
machinery of finite-gap integration method.  In this section we
propose a simple heuristic approach that enables one to link the
spectral DOS $f(\lambda)$ (or $f_\pm(\lambda)$) of a soliton gas with
the ensemble averages of conserved quantities of the integrable
system~\eqref{eq:disp_Euler}. As an illustration we consider the three
first conserved densities of the Euler system~\eqref{eq:disp_Euler}:
$\rho$, $u$ and $\rho u$.

We first consider a homogeneous soliton gas, i.e. a gas whose
statistical properties, particularly the DOS, do not depend on
$x,t$. The proposed approach is based on the natural assumption that
the nonlinear wave field in a homogeneous soliton gas represents an
ergodic random process, both in $x$ and $t$ (we note in passing that
ergodicity is inherent in the model of soliton gas based on the
finite-gap theory, see e.g.~\cite{flaschka_multiphase_1980,
moser_integrable_1981,shurgalina_nonlinear_2016}). The ergodicity
property implies that ensemble-averages $\langle \rho(x,t) \rangle$,
$\langle u(x,t) \rangle$ and $\langle \rho(x,t) u(x,t) \rangle$ in the
soliton gas can be replaced by the corresponding spatial
averages. Generally, for any functional $H[\rho(x,t), u(x,t)] $ we
have
\begin{equation}
\label{eq:psi2b_def}
\langle H[\rho, u] \rangle =  \lim \limits_{L \to \infty} \frac{1}{2L}
\int\limits_{x-L}^{x+L} H[\rho(y,t), u(y,t)] dy,
\end{equation}
for a single representative realization of soliton gas.  We detail
below the derivation of $\langle \rho \rangle $, the generalization to
$\langle u \rangle$ and $\langle \rho u \rangle$ being
straightforward.

Let the soliton gas propagate on a constant background
$(\rho, u)=(\rho_0, u_0)$ (without loss generality one can assume
$(\rho_0, u_0)=(1,0)$).  Let
$\langle \rho \rangle = \rho_0 + \langle \eta \rangle$ where
$\eta = \rho - \rho_0$.  We consider the general, anisotropic case for
which the soliton gas is characterized by two DOS's $f_-(\lambda)$ and
$f_+(\lambda)$.  Define
\begin{equation}\label{cons_I}
I = \int\limits_{x-L}^{x+L} \eta(y,t) dy,
\end{equation}
where $L \gg 1$. Then
$\langle \rho \rangle = \rho_0 + I/(2L) + \mathcal{O}(L^{-1})$.
\begin{figure}[h]
\centering
\includegraphics{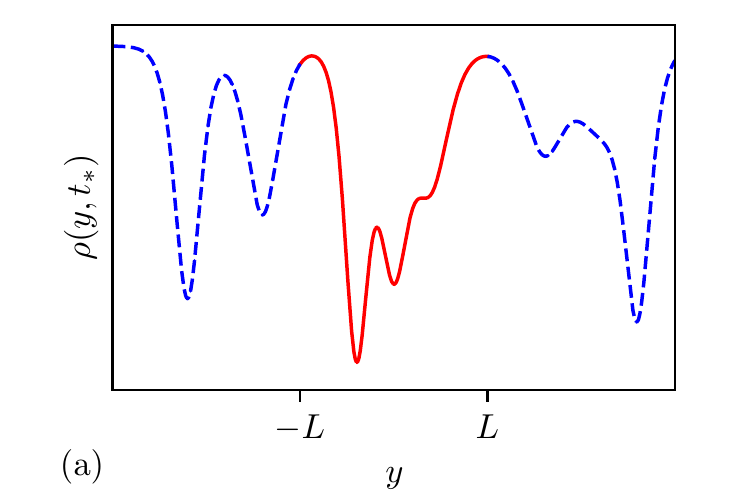}
\includegraphics{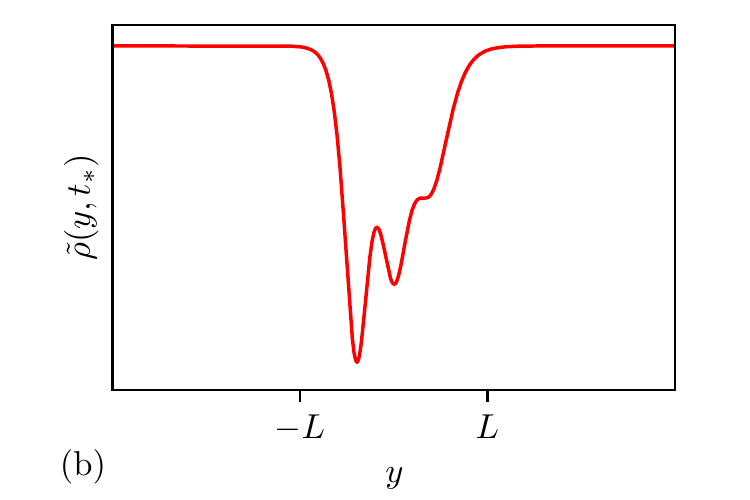}
\includegraphics{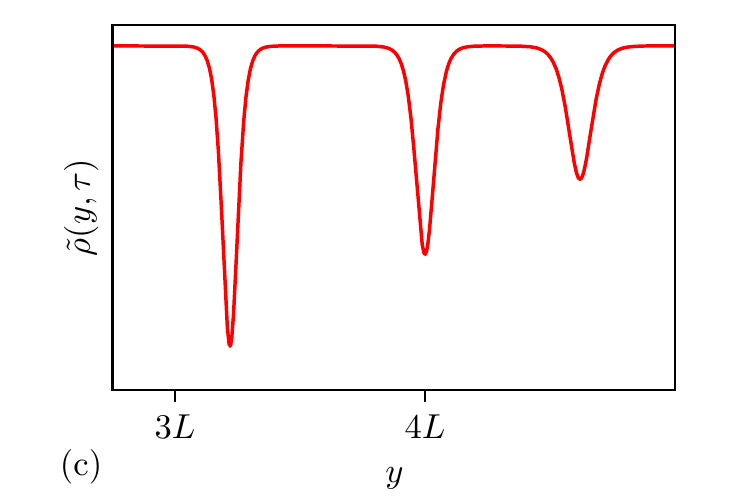}
\caption{Schematic for the evaluation of the integral~\eqref{cons_I}
in soliton gas using the truncation procedure. (a)~Typical
distribution $\rho(y,t_*)$ for a DNLS dark soliton gas. (b)~Truncation
of the distribution $\rho(y,t_*)$ for $y \in (-L,L)$. (c)~Variation of
the truncated distribution $\tilde \rho(y,\tau)$ at time
$\tau \gg t_*$.}
\label{fig:trunc}
\end{figure}

Let $(\rho(y,t),u(y,t))$ be a realization of a soliton gas solution to
the dispersive hydrodynamics~\eqref{eq:disp_Euler} and let
$(\tilde \rho(y,t),\tilde u(y,t))$ be defined in such a way that for
some $t=t_*$ one has
$(\tilde \rho(y,t_*),\tilde u(y,t_*)) = (\rho(y,t_*),u(y,t_*))$ for
$y \in (x-L, x+L)$ and
$(\tilde \rho(y,t_*),\tilde u(y,t_*))=(\rho_0,0)$ outside of this
interval. To avoid complications we assume that the transition between
the two behaviors is smooth but sufficiently rapid so that such a
``windowed'' portion of a soliton gas (see
Fig.~\ref{fig:trunc}) can be approximated by
$N$-soliton solution of~\eqref{eq:disp_Euler} for some $N \gg 1$, with
the discrete IST spectrum being distributed on $\Omega_-$ and $\Omega_+$ with densities
$2Lf_- (\lambda)$ and $2Lf_+ (\lambda)$ 
respectively (recall the definition of DOS in Sec.~\ref{sec:intro}).
Equation~\eqref{cons_I} rewrites
\begin{equation}
\label{eq:IJ}
I = \int\limits_{-\infty}^{+\infty}  \tilde \eta(y,t)
dy,\quad
\tilde \eta(y,t) = \tilde \rho(y,t) -\rho_0.
\end{equation}
We note the integral~\eqref{eq:IJ} does not depend on time because $I$
is a conserved quantity, in particular, for $t=\tau \gg t_*$ where the
solution $(\tilde \rho(y,\tau),\tilde u(y,\tau))$ asymptotically
represents the train of spatially well-separated solitons
$\rho_{\rm s}^\pm, u_{\rm s}^\pm$ propagating on the background
$(\rho_0, 0)$ (see Fig.~\ref{fig:trunc}). In this case, $I$ can be
evaluated as
\begin{align}
\label{eq:sum_I}
I = &\sum_i \int\limits_{-\infty}^{+\infty}
(\rho^-_{\rm s}(y-\lambda_i \tau- y_i;\lambda_i)-\rho_0) dy\notag\\
&+ \sum_j \int
\limits_{-\infty}^{+\infty}
(\rho^+_{\rm s}(y-\lambda_j \tau- y_j;\lambda_j)-\rho_0) ,
\end{align}
where $\lambda_{i,j}$ are the spectral parameters and $y_{i,j}$ the
initial phases of the $\pm$-solitons. Since the spectrum is preserved
by the integrable dynamics~\eqref{eq:disp_Euler}, $\lambda_{i,j}$
remain to be distributed on $\Omega_{\pm}$ with the  respective densities  $2L f_\pm(\lambda)$ for all
$t$. Let $\overline{\eta}_\pm$ be the ``mass'' of the spectral soliton
solution $\rho_{\rm s}^{\pm}(x-\lambda t; \lambda)-\rho_0$,
\begin{equation}
\label{eq:rhob_s}
\overline{\eta}_\pm(\lambda) =\int\limits_{-\infty}^{+\infty}
(\rho^\pm_{\rm s} (y;\lambda)-\rho_0) dy,
\end{equation}
which only depends on $\lambda$. Note that the integral
in~\eqref{eq:rhob_s} converges for the example considered in
Sec.~\ref{sec:ex} since $\rho^\pm_{\rm s}$ decays exponentially to
$\rho_0$.  We have with this new notation:
$I = \sum_i \overline{\eta}_-(\lambda_i) + \sum_j
\overline{\eta}_+(\lambda_j)$.  Taking the continuous limit,
$\sum_i \to \int_{\Omega_-} d\lambda \, 2L f_-(\lambda)$ and
$\sum_j \to \int_{\Omega_+} d\lambda \,2L f_+(\lambda)$, we obtain:
\begin{equation}
\frac{I}{2L} = \int_{\Omega_-} \overline{\eta}_-(\lambda)
f_-(\lambda) d\lambda + \int_{\Omega_+}
\overline{\eta}_+(\lambda)
f_+(\lambda) d\lambda,
\end{equation}
yielding the expression for the moment $\langle \rho \rangle$:
\begin{align}
\label{eq:I}
\langle \rho(x,t)\rangle = \rho_0 + \int_{\Omega_-}
\overline{\eta}_-(\lambda) f_-(\lambda) d\lambda \notag\\
+ \int_{\Omega_+} \overline{\eta}_+(\lambda) f_+(\lambda) d\lambda.
\end{align}
Similarly, we obtain for the two other moments (recall that we assume
$u \to 0$ as $x \pm \infty$):
\begin{align}
\label{eq:I2}
\langle u(x,t)\rangle =  &\int_{\Omega_-}
\overline{u}_-(\lambda) f_-(\lambda) d\lambda \notag\\
&+ \int_{\Omega_+} \overline{u}_+(\lambda) f_+(\lambda,x,t) d\lambda,
\\
\label{eq:I3}
\langle \rho(x,t)u(x,t)\rangle =  &\int_{\Omega_-}
\overline{\rho u}_-(\lambda) f_-(\lambda) d\lambda \notag\\
&+
\int_{\Omega_+} \overline{\rho u}_+(\lambda) f_+(\lambda)
d\lambda,
\end{align}
where $\overline{u}_\pm(\lambda) = \int u_{\rm s}^\pm(y;\lambda) dy$
and
$\overline{\rho u}_\pm(\lambda) = \int u_{\rm s}^\pm(y;\lambda) u_{\rm
s}^\pm(y;\lambda) dy$. The expressions~\eqref{eq:I}, \eqref{eq:I2}
and~\eqref{eq:I3} rewrite in the isotropic case:
\begin{equation}
\label{eq:I4}
\begin{split}
\langle \rho(x,t)\rangle &= \rho_0 + \int_{\Omega}
\overline{\eta}(\lambda) f(\lambda) d\lambda,\\
\langle u(x,t)\rangle &=  \int_{\Omega}
\overline{u}(\lambda) f(\lambda) d\lambda,\\
\langle \rho(x,t) u(x,t)\rangle &=  \int_{\Omega}
\overline{\rho u}(\lambda) f(\lambda) d\lambda.
\end{split}
\end{equation}
We present in Table~\ref{tab:mass} the expressions of
$\overline{\eta}_\pm(\lambda)$, $\overline{u}_\pm(\lambda)$ and
$\overline{\eta}_\pm(\lambda)$ for the examples introduced in
Sec.~\ref{sec:ex}.
\begin{table*}
\centering
\begin{tabular}{p{3.5cm} p{3.5cm} p{3.5cm} p{3.5cm}}
\hline\hline
equations & $\overline{\eta}(\lambda)$ & $\overline{u}(\lambda)$ &
$\overline{\rho u}(\lambda)$  \\
\hline\hline
DNLS ($\sigma=+1$)&$-2\sqrt{1-\lambda^2}$ & $2 \sin^{-1}(\lambda)-\pi
\sgn (\lambda)$&
$-2 \lambda \sqrt {1 -\lambda^2}$\\ 
RNLS ($\sigma=-1$)&$+2\sqrt{\lambda^2-1}$ & $2 \sgn(\lambda)
\cosh^{-1} |\lambda|$ & $+2 \lambda \sqrt {\lambda^2-1}$\\
\hline
\end{tabular}
\caption{Expressions of the integral $\overline{\eta}_\pm(\lambda)$,
$\overline{u}_\pm(\lambda)$ and $\overline{\eta}_\pm(\lambda)$ for NLS
solitons ($\rho_0=1$). For both examples we have
$\rho_{\rm s}^-=\rho_{\rm s}^+$ and $u_{\rm s}^-=u_{\rm s}^+$ such
that $\overline{\eta}_- = \overline{\eta}_+$,
$\overline{u}_- = \overline{u}_+$ and
$\overline{\rho u}_- = \overline{\rho u}_+$.}
\label{tab:mass}
\end{table*}

The method presented here only requires to integrate the
single-soliton solution and thus can be readily applied to any
integrable dispersive hydrodynamic system supporting the soliton
resolution scenario. Formulas~\eqref{eq:I}, \eqref{eq:I2},
\eqref{eq:I3} and~\eqref{eq:I4} will be used in the next section to
track the evolution of the DOS numerically.  In conclusion we note
that the above simple method, applied to the KdV equation, gives
exactly the same results for the mean and mean square of the random
field as the finite-gap theory consideration
of~\cite{el_thermodynamic_2003, el_critical_2016}.  It also explains
why the corresponding analytical expressions for the moments in a
dense gas of KdV solitons derived in~\cite{el_critical_2016} coincide
with the corresponding expressions obtained
in~\cite{dutykh_numerical_2014} for a rarefied gas (see
also~\cite{shurgalina_nonlinear_2016} for the similar modified KdV
equation result).

In the above consideration of homogeneous soliton gases the ensemble
averages~\eqref{eq:psi2b_def} are constant. For a nonhomogeneous gas
the DOS is a slowly varying function of $x,t$ and so are the ensemble
averages that now need to be interpreted as ``local averages'' in the
spirit of modulation theory~\cite{whitham_linear_1999}.  Essentially,
one introduces a mesoscopic scale $\ell$, much larger than the
typical soliton width and much smaller than the spatial scale of the
DOS variations so that the DOS is approximately constant on any
interval $(x-\ell, x+\ell)$. Then the constant ensemble averages
\eqref{eq:psi2b_def} are replaced by slowly varying quantities:
\begin{equation}
\label{eq:ell}
\langle H[\rho, u] \rangle_\ell (x,t)= \frac{1}{2\ell} \int
\limits_{x-\ell}^{x+\ell} H[\rho(y,t), u(y,t)] dy.
\end{equation}
The local averages $\langle H[\rho, u] \rangle_\ell$ do not depend on
$\ell$ at leading order, and their spatiotemporal variations occur on
$x,t$-scales that correspond to the scales associated with variations
of $f(\lambda)$ and are much larger than those of $\rho$, $u$. The
modulations of $\langle \rho \rangle$, $\langle u \rangle$ and
$\langle u \rangle$ in a nonhomogeneous soliton gas are then defined
by the equations~\eqref{eq:I}, \eqref{eq:I2} and~\eqref{eq:I3}
respectively, in which the DOS $f_\pm(\lambda)$ is replaced by the
solution $f_\pm(\lambda, x, t)$ of the kinetic
equation~\eqref{eq:kin2},\eqref{eq:s2}. This strategy will be used in
the next section where we study dynamics of nonhomogeneous soliton
gases generated in the solutions of Riemann problems for kinetic
equations.

\section{Multi-component bidirectional soliton gases:
Riemann problem}
\label{sec:beam}

\subsection{Hydrodynamics reductions}

Generally, our ability to solve the integral equation of
state~\eqref{eq:state} is very limited, and strongly depends on the
particular form of the interaction kernel.  Some particular analytical
solutions have been found~\cite{el_spectral_2020} for special cases of
soliton gases for the focusing NLS equation.  At the same time, it was
shown
in~\cite{el_kinetic_2005,el_kinetic_2011,pavlov_generalized_2012} that
this problem greatly simplifies if discretization the DOS
$f(\lambda,x,t)$ or $f_\pm(\lambda,x,t)$ with respect to the soliton
spectral parameter~$\lambda$ is admissible.  We adopt this
simplification in the following, and we consider the soliton gases
that are composed of a finite number of distinct spectral components,
termed monochromatic, or cold, components.  We consider in the
following the general anisotropic description; the derivation also
readily applies to the isotropic case. Suppose that the bidirectional
soliton gas is spectrally composed of $n_-$ distinct components of the
``$-$'' soliton branch, and $n_+$ distinct components of the ``$+$''
soliton branch:
\begin{equation}
\label{eq:F0}
\begin{split}
&f_-(\lambda,x,t) = \sum_{i=1}^{n_-} F_i(x,t)
\delta(\lambda-\Lambda_i),\\
&f_+(\lambda,x,t) = \sum_{i=n_-+1}^{n_-+n_+} F_i(x,t)
\delta(\lambda-\Lambda_i),
\end{split}
\end{equation}
with $c_\pm(\Lambda_{i})<c_\pm(\Lambda_{i+1})$ and where $\Lambda_i$ are the soliton parameters of the different
components and $\delta$ the Dirac delta distribution.  We do not
indicate in the following the branch-belonging of the component $F_i$
for readability reason. Additionally, we do not indicate explicitly
the $(x,t)$-dependence of the fields $F_i$ when it is clear.  As
pointed out in~\cite{carbone_macroscopic_2016, el_spectral_2020}, the
multi-component ansatz~\eqref{eq:F0} is a mathematical idealization,
physically one would replace the $\delta$-functions by narrow
distributions around the spectral points $\Lambda_i$.

The ansatz~\eqref{eq:F0} transforms the pair of distributions
$(f_-(\lambda),f_+(\lambda))$ into a $n=n_-+n_+$-dimensional vector
$\bs F = (F_1, \dots, F_n)$. Thus~\eqref{eq:kin2} reduces to $n$
hydrodynamic (quasi-linear) conservation laws:
\begin{equation}
\label{eq:F_NLS}
(F_i)_t + (S_i F_i)_x = 0,\quad i =1\dots n,
\end{equation}
where $S_i(x,t) = s_{\pm_i}(\Lambda_i,x,t)$ with $\pm_i$ indicating
the branch-belonging of the soliton $\Lambda_i$.  The coupled
equations of states~\eqref{eq:s2} simplify into a $n$th-order linear
algebraic system for the $S_i$'s:
\begin{equation}
\label{eq:S_NLS0}
S_i  =  C_i + \sum_{j \neq i} \Delta(\Lambda_i,\Lambda_j)
F_j |S_i-S_j|,\quad C_i = c_{\pm_i}(\Lambda_i).
\end{equation}
The system~\eqref{eq:S_NLS0} simplifies for the examples considered in
Sec.~\ref{sec:ex}, where we assumed that $S_{i} < S_{i+1}$ and where
the phase shift formula has the form:
$\Delta_{\pm\pm} (\lambda, \mu)= \Delta_{\pm\mp} (\lambda, \mu) =
\sgn(\lambda-\mu) G(\lambda, \mu)$; the expression of $G$ is given
by~\eqref{eq:Delta_DNLS} for the DNLS and~\eqref{eq:DeltaRNLS} for the
RNLS equation. For the NLS examples we obtain the linear system:
\begin{equation}
\label{eq:S_NLS}
S_i  =  C_i + \sum_{j \neq i} G_{ij}
F_j (S_i-S_j),\quad G_{ij} = G(\Lambda_i,\Lambda_j).
\end{equation}
In order to simplify the discussion, we will focus on the latter
system. Both anisotropic and isotropic soliton gases are described by
the same system~\eqref{eq:F_NLS},\eqref{eq:S_NLS}: for the isotropic
DNLS-soliton gas we have $G_{ij}=G_1(\Lambda_i,\Lambda_j)>0$, and for
the anisotropic RNLS-soliton gas
$G_{ij}=G_2(\Lambda_i,\Lambda_j) \in \mathbb{R}$.

The resolution of the linear
system~\eqref{eq:S_NLS} yields a solution $S_i(\bs{F})$ such that the
system~\eqref{eq:F_NLS} becomes quasi-linear:
\begin{equation}
\label{eq:F}
(F_i)_t + (S_i(\bs F) F_i)_x = 0.
\end{equation}
It was shown in~\cite{el_kinetic_2011,pavlov_generalized_2012} that
the system~\eqref{eq:F} is a linearly degenerate integrable
system~\cite{ferapontov_integration_1991} and its general solutions
can be obtained using the generalized hodograph
method~\cite{tsarev_geometry_1991}. In particular, the characteristic
velocities of this hydrodynamic system coincide with the mean
velocities $S_i$.

Finally, the expressions of the moments $\langle \rho \rangle$,
$\langle u \rangle$ and $\langle \rho u \rangle$ are given by:
\begin{equation}
\label{eq:mi}
\begin{split}
\langle \rho(x,t)\rangle &= \rho_0 + \sum_{i=1}^n
\overline{\eta}(\Lambda_i) F_i(x,t) ,\\
\langle u(x,t)\rangle &=  \sum_{i=1}^n
\overline{u}(\Lambda_i) F_i(x,t),\\
\langle \rho(x,t) u(x,t)\rangle &=  \sum_{i=1}^n
\overline{\rho u}(\Lambda_i) F_i(x,t),
\end{split}
\end{equation}
with the coefficients $\overline{\eta}$, $\overline{u}$ and
$\overline{\rho u}$ given in Table~\ref{tab:mass} for the NLS
equation~\eqref{eq:NLS}. The relations in~\eqref{eq:mi} can be used to
obtain the DOS components $F_i$ from the moments
$\langle \rho \rangle$, $\langle u \rangle$ and
$\langle \rho u \rangle$ if $n \leq 3$.

\subsection{Shock-tube problem}
\label{sec:riem}

We now focus on the physically relevant Riemann problem for the
hydrodynamic system~\eqref{eq:S_NLS},\eqref{eq:F}  describing the
interaction dynamics of two soliton gases prepared in the respective
uniform states $\bs F^{\rm L} \in \mathbb{R}^n$ and
$\bs F^{\rm R}\in \mathbb{R}^n$, that are initially separated:
\begin{equation}
\label{eq:init}
\bs F(x,0) =
\begin{cases}
\bs F^{\rm L}, &\text{if } x<0,\\
\bs F^{\rm R}, &\text{if } x\geq 0.
\end{cases}
\end{equation}
The spectral distribution~\eqref{eq:init} corresponds to the soliton
gas ``shock tube'' problem, an analog of the standard shock tube
problem of classical gas dynamics.  The shock tube problem represents
a good benchmark for our kinetic theory where we can investigate both
overtaking and head-on collisions by choosing the appropriate number
of components. We emphasize here that the initial
condition~\eqref{eq:init} constitutes a Riemann problem for the
kinetic equation~\eqref{eq:F} but not for the original dispersive
hydrodynamics system~\eqref{eq:disp_Euler}, similar to the so-called
generalized Riemann problems recently introduced
in~\cite{sprenger_discontinuous_2020,gavrilyuk_stationary_2020}. We
shall sometimes refer to the problem~\eqref{eq:F},\eqref{eq:init} as a
``spectral Riemann problem'' as it essentially describes the
spatiotemporal evolution of the spectral components of the soliton
gas.

The soliton gas shock tube problem has been investigated for the KdV
and focusing NLS two-component soliton gases ($n=2$)
in~\cite{el_kinetic_2005, carbone_macroscopic_2016, el_spectral_2020},
and for $n$ components in the context of generalized
hydrodynamics~\cite{castro-alvaredo_emergent_2016,
bertini_transport_2016, doyon_dynamics_2017,
kuniba_generalized_2020,croydon_generalized_2020}. Here we present the
problem for the $n$-component bidirectional anisotropic soliton
gases. An important difference of our consideration from the
generalized hydrodynamics setting is that we are interested not only
in the spectral characterization of soliton gases via solutions of the
kinetic equations but also (and ultimately) in the description of the
classical nonlinear wave fields associated with these solutions. The
latter is achieved by the evaluation of the ensemble averages as
described in Section~\ref{sec:mean}.

Due to the scaling invariance of the problem (the kinetic
equation~\eqref{eq:F} and the initial condition~\eqref{eq:init} are
both invariant with respect to the transformation $x \to Cx$,
$t \to Ct$), the solution is a self-similar distribution $\bs
F(x/t)$. Because of the linear degeneracy of the quasi-linear
system~\eqref{eq:F} the only admissible solutions are constant
separated by contact discontinuities, cf. for
instance~\cite{rozhdestvenskii_systems_1983}. Discontinuous, weak,
solutions are physically acceptable here since the kinetic equation
describes the conservation of the number of solitons within any given
spectral interval, and Rankine-Hugoniot type conditions can be imposed
to ensure the conservation of the number of solitons across
discontinuities. The solution of the Riemann problem is composed of
$n+1$ constant states, or plateaus, separated by $n$ discontinuities
(see e.g.~\cite{lax_hyperbolic_1973}):
\begin{equation}
\label{eq:sol}
F_i(x,t) =
\begin{cases}
F_i^{1} = F_i^{\rm L}, &x/t<Z_1,\\
\dots\\
F_i^j, & Z_{j-1} \leq x/t < Z_j\\
\dots\\
F_i^{n+1} = F_i^{\rm R}, &Z_n \leq x/t,
\end{cases}\quad
\end{equation}
where the index $i$ indicates the $i$-th component of the vector
$\bs{F}$, and the exponent $j$ the
index of the plateau. For clarity we labeled the superscripts $j=1$ as
``L'' (left boundary condition) and $j=n+1$ as ``R'' (right boundary
condition).  Additionally the index $j$ of the plateau's value
$F_i^j$ will be written as a Roman numeral in the examples considered
later on.  The contact discontinuities propagate at the characteristic
velocities~\cite{lax_hyperbolic_1973}:
\begin{equation}
\label{eq:z}
Z_j = S_{j}(F_1^j,\dots,F_n^j) = S_{j}(F_1^{j+1},\dots,F_n^{j+1}),
\end{equation}
where the plateaus' values $F_i^j$ are given
by Rankine-Hugoniot jump conditions:
\begin{align}
\label{eq:RH}
-Z_j \Big[ F_i^{j+1} -F_i^{j}\Big] + \Big[
&S_{i}(F_1^{j+1},\dots,F_n^{j+1})
F_i^{j+1} \\  &-  S_{i}(F_1^j,\dots,F_n^j)
F_i^{j} \Big] = 0,
\end{align}
where $i,j=1 \dots n$.
The Rankine-Hugoniot conditions with $i=j$ are trivially satisfied by
the definition of contact discontinuity~\eqref{eq:z}. Recalling the
effective derivation of the equation of state in
Sec.~\ref{sec:general}, the velocity of the contact discontinuity
$Z_j$ can be identified as the velocity of a trial soliton with
parameter $\Lambda_j$ propagating in a soliton gas of density
$\bs{F} = (F_1^j, \dots F_n^j)$ or equivalently
$\bs{F} = (F_1^{j+1}, \dots F_n^{j+1})$.

Note that, if the solitons were not interacting, the initial step
distribution $F_i(x,0)$ for the component $\lambda = \Lambda_i$ would
have propagated at the free soliton velocity $C_i$:
\begin{equation}
\label{eq:sol_free}
F^{\rm free}_i(x,t) =
\begin{cases}
F_i^{\rm L}, &x/t< C_i,\\
F_i^{\rm R}, &C_i \leq x/t,
\end{cases}
\quad i=1 \dots n,
\end{equation}
which dramatically differs from the solution~\eqref{eq:sol}.  In order
to demonstrate the validity of the
solution~\eqref{eq:sol},\eqref{eq:z},\eqref{eq:RH} the Riemann
problem is investigated numerically for the DNLS and RNLS equations
for two- and three-component soliton gases in the next sections.

\subsubsection{Two-component soliton gas}
\label{sec:ex1}

We consider in this section the interaction between two components of
soliton gas with respective parameters $\Lambda_1$ and $\Lambda_2$
(recall that $S_1 < S_2$).  The solution of the equation of
state~\eqref{eq:S_NLS} reads for $n=2$:
\begin{equation}
\label{eq:S12}
\begin{split}
S_1(F_1,F_2) &= \frac{(1- G_{21} F_1) C_1-G_{12} F_2 \, C_2 }{1- G_{21}
F_1 - G_{12} F_2},\\
S_2(F_1,F_2) &= \frac{(1- G_{12} F_2) C_2-G_{21} F_1 \, C_1 }{1- G_{21} F_1 -
G_{12} F_2 }.
\end{split}
\end{equation}
As noted in~\cite{el_kinetic_2005}, the densities $F_1$ and $F_2$ must
satisfy the inequality:
\begin{equation}
\label{eq:det1}
G_{21} F_1 + G_{12} F_2 < 1,
\end{equation}
for the expressions~\eqref{eq:S12} to remain valid; we suppose that
this condition is always verified, constraining the DOS in the
following.  We suppose that $F_1^{\rm L}=F_2^{\rm R}=0$ and
$F_1^{\rm R}=F_2^{\rm L}=\zeta_0$: the region $x<0$ is initially only
populated with $\Lambda_2$-solitons and the region $x>0$ of slower
$\Lambda_1$-solitons. Since $S_1 < S_2$ the two ``species'' of soliton
are interacting.  Note that~\eqref{eq:det1} implies
$G_{12} \zeta_0,G_{21} \zeta_0 <1$.  The solution~\eqref{eq:sol} has 3
plateaus:
\begin{equation}
\label{eq:sol21}
F_i(x,t) =
\begin{cases}
F_i^{\rm I} = \delta_{i,2} \zeta_0, &x/t<Z_1,\\
F_i^{\rm II}, & Z_1 \leq x/t < Z_2,\\
F_i^{\rm III} = \delta_{i,1} \zeta_0, &Z_2 \leq x/t,
\end{cases}\quad
\end{equation}
where $i \in \{1,2\}$, 
with the value at the intermediate plateau:
\begin{equation}
\label{eq:sol22}
F_1^{\rm II} = \frac{[1-G_{12} \zeta_0] \zeta_0}
{1-G_{12} G_{21}\zeta_0^2},\quad
F_2^{\rm II} = \frac{[1-G_{21} \zeta_0]\zeta_0}
{1-G_{12} G_{21}\zeta_0^2},
\end{equation}
and the velocities of the discontinuities:
\begin{equation}
\label{eq:sol23}
\begin{split}
&Z_1 = S_1(0,\zeta_0) = \frac{
C_1-G_{12} \zeta_0 \, C_2 }{1-
G_{12} \zeta_0},\\
&Z_2 = S_2(\zeta_0,0) = \frac{
C_2-G_{21} \zeta_0 \, C_1 }{1-
G_{21} \zeta_0}.
\end{split}
\end{equation}

Both kinds of soliton propagate in the region delimited by $x= Z_1 t$
and $x= Z_2 t$ (since $F_1^{\rm II} \neq 0$, $F_2^{\rm II} \neq 0$),
and we refer to this region as the interaction region  in the
following. The discontinuity's velocity $Z_i$ corresponds to the
effective velocity of solitons $\Lambda_i$ in this region.  The total
density of solitons $\sum_i F_i$ in the interaction region  is
given by:
\begin{equation}
\label{eq:solt}
F_1^{\rm II} + F_2^{\rm II} = \frac{2-(G_{12}
+G_{21}) \zeta_0}
{1-G_{12} G_{21} \zeta_0^2} \,
\zeta_0.
\end{equation}
If $\sgn(G_{12})=\sgn(G_{21})>0$ ($<0$) then the total density
$F_1^{\rm II} + F_2^{\rm II}$ is smaller (larger) than the sum of the
initial soliton densities $2\zeta_0$, and $Z_1<C_1<C_2<Z_2$
($C_1<Z_1<Z_2<C_2$), cf. for instance~\cite{el_kinetic_2005}.

The two-component shock tube problem ($n=2$) has been 
investigated numerically in~\cite{carbone_macroscopic_2016} for KdV
soliton gases. We have shown in Sec.~\ref{sec:ex} that the kinetic
dynamics of the KdV soliton and the isotropic DNLS soliton gas are
both governed by equations~\eqref{eq:conserv},\eqref{eq:s_DNLS} with
$G_1(\lambda,\mu)>0$. Thus solutions of the DNLS spectral Riemann
problem and the KdV spectral Riemann problem are expected to describe
very similar dynamics, and we rather focus on the anisotropic RNLS
soliton gas exhibiting two distinct kinds of interaction. The solution
of the RNLS spectral Riemann problem is given
by~\eqref{eq:sol21},\eqref{eq:sol22},\eqref{eq:sol23} where
$G_{ij}=G_2(\Lambda_i,\Lambda_j)$ with $G_2$ defined
in~\eqref{eq:DeltaRNLS}.

To verify the validity of our spectral solutions in the context of the
original nonlinear wave problem of the interaction of soliton gases,
we solve numerically the RNLS equation~\eqref{eq:RNLS} with initial
conditions corresponding to the spectral Riemann data~\eqref{eq:init}
for two different RNLS soliton gases with: (i) overtaking collisions
$G_{ij}>0$ ($\Lambda_1=1.05,\Lambda_2 \in [1.06,1.1]$), and (ii)
head-on collisions $G_{ij}<0$
($\Lambda_1=-1.05,\Lambda_2 \in [1.06,1.1]$). The boundary values
$\bs F^{\rm L}$ and $\bs F^{\rm R}$ for cases (i) and (ii) are
indicated in Table~\ref{tab:num}.  $50$ initial conditions
$\rho(x,0),u(x,0)$ are realized according to the initial step
distribution~\eqref{eq:init} and evolved through a direct numerical
simulation of the NLS equation~\eqref{eq:NLS} with $\sigma=-1$.  The
details of the numerical implementation of the initial
condition~\eqref{eq:init} and the direct numerical resolution
of~\eqref{eq:NLS} are given in Appendix~\ref{sec:num}.
\begin{table*}
\centering
\begin{tabular}{p{1cm} p{5cm} p{4cm} p{4cm}}
\hline\hline
& soliton parameter & left boundary condition & right boundary
condition\\
\hline\hline
(i) & $(\Lambda_1 = 1.05 ,\Lambda_2 \in [1.06,1.10])$ &
$\bs F^{\rm L}=(0,6.6) \times 10^{-2}$ & $\bs F^{\rm R}=(6.6,0) \times
10^{-2}$ \\ 
(ii) & $(\Lambda_1 = -1.05,\Lambda_2 \in [1.06,1.1])$ &
$\bs F^{\rm L}=(0,6.6) \times 10^{-2}$ & $\bs F^{\rm R}=(6.6,0) \times
10^{-2}$ \\ 
(iii) & $(\Lambda_1,\Lambda_2,\Lambda_3)=(-0.2,0.1,0.4)$ &
$\bs F^{\rm L}=(2.5,0,7.5) \times 10^{-2}$ & $\bs F^{\rm R}=(5,5,0)
\times 10^{-2}$ \\ 
(iv) & $(\Lambda_1,\Lambda_2,\Lambda_3)=(-1.1,1.05,1.1)$ &
$\bs F^{\rm L}=(1.6,0,5) \times 10^{-2}$ & $\bs F^{\rm R}=(3.3,3.3,0)
\times 10^{-2}$ \\ 
\hline
\end{tabular}
\caption{Initial conditions for the spectral Riemann
problem~\eqref{eq:F},\eqref{eq:init} considered in
Sec.~\ref{sec:riem}. The constraint on the spectral parameters
$|\Lambda_i| \leq 1.1$ in (i), (ii) and (iv) is due to the limits of
the numerical scheme used to solve the RNLS equation,
cf. Appendix~\ref{sec:num}. }
\label{tab:num}
\end{table*}

\begin{figure}[h]
\centering
\includegraphics{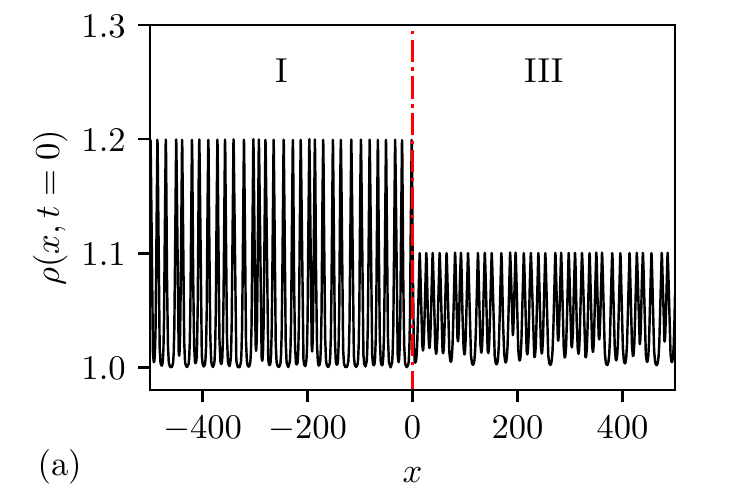}
\includegraphics{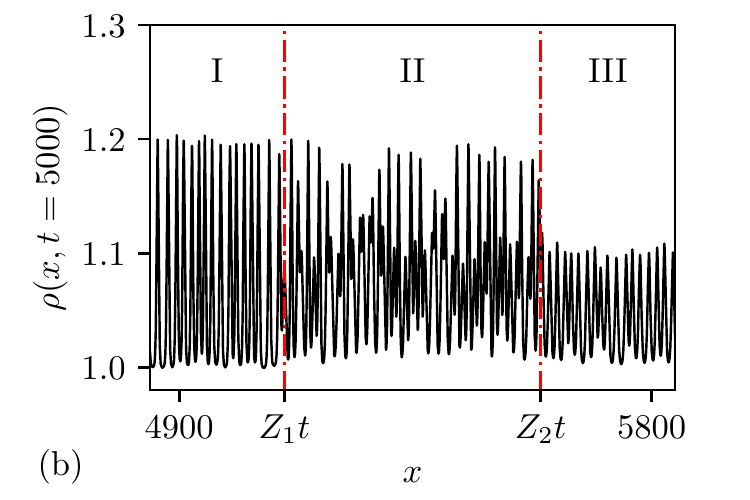}
\caption{Example of one realization of the soliton gas shock tube
problem (i) at $t=0$~(a) and $t=5000$~(b) with
$(\Lambda_1,\Lambda_2) = (1.05,1.10)$. The two regions I and III
corresponds respectively to the left and right boundary conditions
prescribed in the initial condition, cf.~\eqref{eq:init}. The
variation of $\rho(x,t)$ clearly displays the formation of an
intermediate interaction region, denoted region II, between the two
positions $x=Z_1 t$ and $x=Z_2 t$ highlighted by vertical
dash-dotted lines.}
\label{fig:psi2}
\end{figure}

A typical RNLS soliton gas distribution $\rho(x,0)$ and its
corresponding numerical evolution $\rho(x,t)$ are displayed in
Fig.~\ref{fig:psi2} for the spectral Riemann problem (i); soliton gas
realizations for the spectral Riemann problem (ii) have a similar
variation with different velocities $Z_1$ and $Z_2$.  We emphasize
that, although the soliton gas is initially prepared in a rarefied
regime where solitons are spatially well-separated
(cf. Appendix~\ref{sec:num}), the total density of solitons increases
in the interaction region, and a dense soliton gas can be observed in
Fig.~\ref{fig:psi2} for which solitons exhibit significant
overlap.

\begin{figure}[h]
\centering
\includegraphics{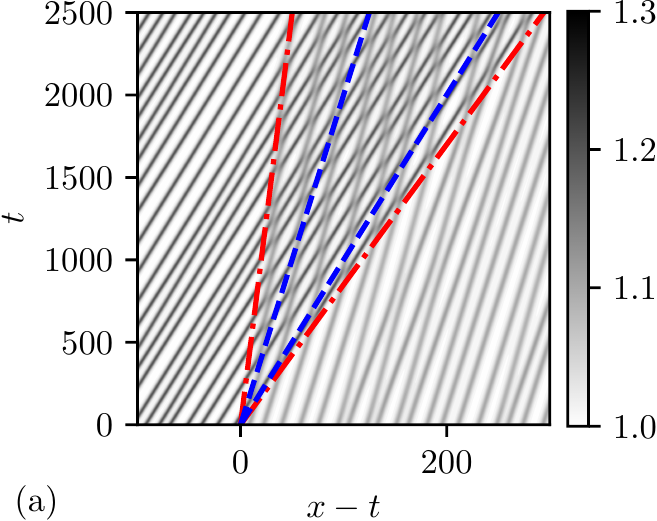}
\includegraphics{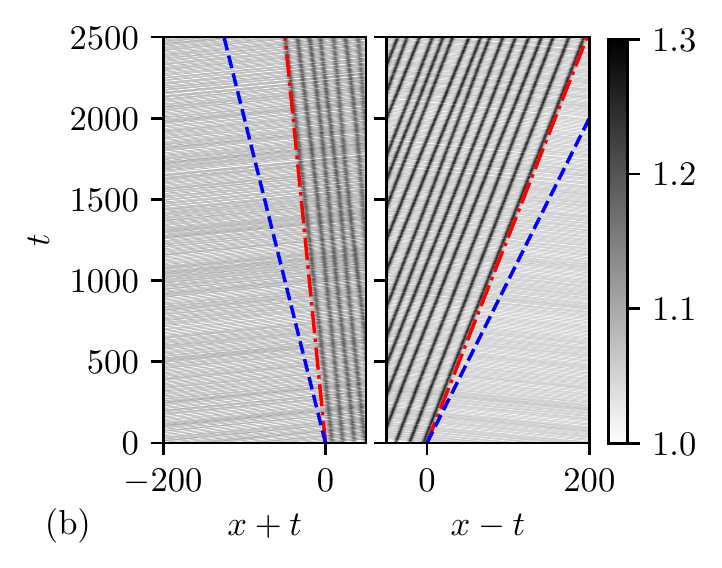}
\caption{Spatiotemporal plots of the field $\rho(x,t)$ for one
realization of the soliton gas. Trajectories of the solitons appear in
solid lines. Dash-dotted lines correspond to the trajectories of the
contact discontinuities: $x = Z_1 t$, $x = Z_2 t$,
cf.~\eqref{eq:sol23}, and dashed lines to the free soliton
trajectories: $x = C_1 t$, $x = C_2 t$. (a)~Overtaking collisions
$(\Lambda_1,\Lambda_2)=(1.05,1.10)$, cf. initial condition (i) in
Table~\ref{tab:num}. (b)~Head-on collisions
$(\Lambda_1,\Lambda_2)=(-1.05,1.10)$, cf. initial condition (ii) in
Table~\ref{tab:num}.}
\label{fig:xt}
\end{figure}

Spatiotemporal evolution of one soliton gas realization is displayed
in Fig.~\ref{fig:xt}, with overtaking collisions (i) and head-on
collisions (ii). To enhance the discrepancy between free soliton
velocities $C_i$ and contact discontinuities velocity $Z_i$ the
trajectories of the solitons are followed in the frames $(x-t,t)$ for
overtaking collisions where $Z_i \sim C_i \sim 1$, and $(x\pm t,t)$
for head-on collisions where $Z_1 \sim C_1 \sim -1$ and
$Z_2 \sim C_2 \sim 1$. One can notice that the interaction time
between two solitons is very short for a head-on collision, which
explains the weakness of head-on interactions compared to overtaking
interactions.


\begin{figure}[h]
\centering
\includegraphics{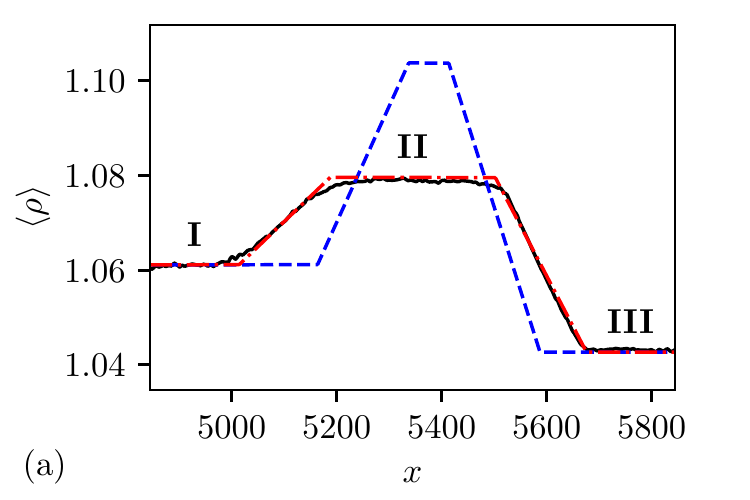}
\includegraphics{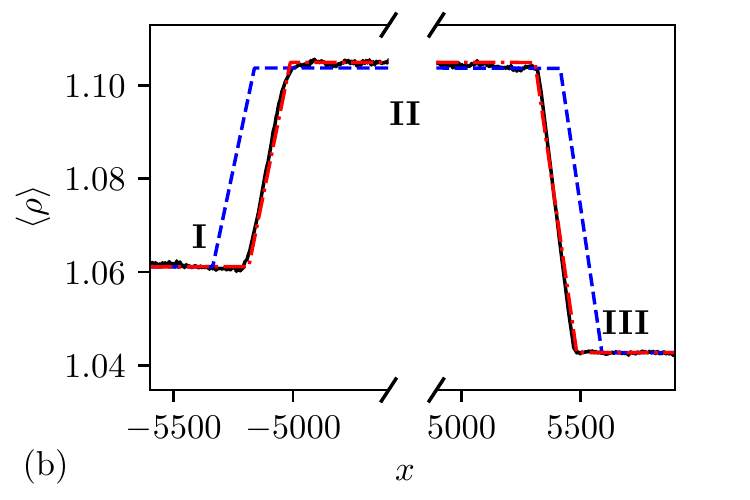}
\caption{Comparison between the ensemble average
$\langle \rho(x,t=5000) \rangle$ of $50$ direct numerical solutions of
the RNLS soliton gas shock tube problem (solid line) and the
analytical solution~\eqref{eq:mi},\eqref{eq:sol21} obtained via the
spectral kinetic theory (dash-dotted line). The dashed lines
correspond to the respective spectral
solutions~\eqref{eq:mi},\eqref{eq:sol_free} for a non-interacting
soliton gas. (a)~Overtaking collisions
$(\Lambda_1,\Lambda_2)=(1.05,1.10)$. (b)~Head-on collisions
$(\Lambda_1,\Lambda_2)=(-1.05,1.10)$.}
\label{fig:rsol2}
\end{figure}

\begin{figure*}
\centering
\includegraphics{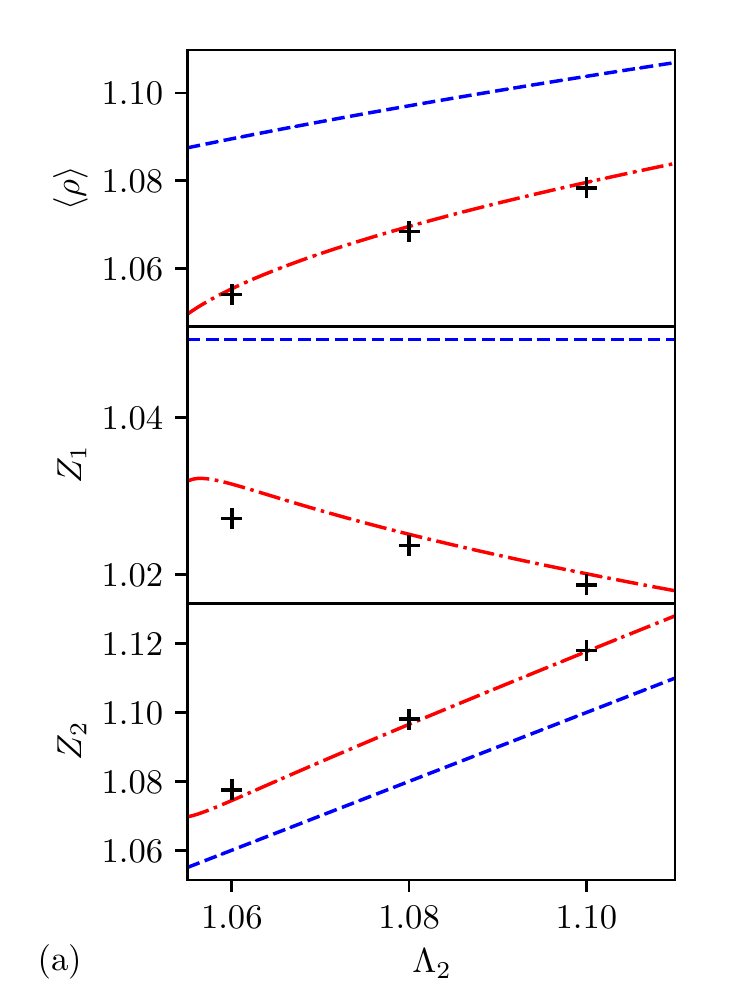}
\includegraphics{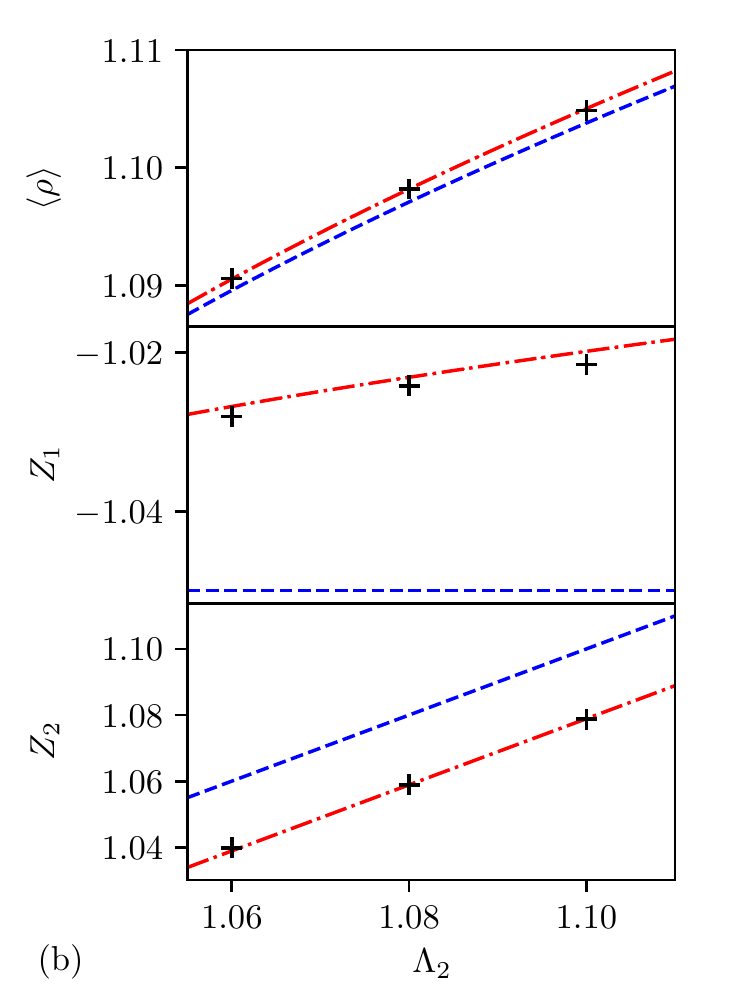}
\caption{Comparison between the parameters of the analytical
solutions~\eqref{eq:mi},\eqref{eq:sol21} (dash-dotted line) and the
corresponding fitted parameters of the numerical solution (crosses)
with different spectral parameters $\Lambda_2$; numerical averages are
obtained over $50$ realizations. For comparison, the dashed lines
correspond to the parameters of the non-interacting solitons
solution~\eqref{eq:mi},\eqref{eq:sol_free}. $\langle \rho \rangle$ is
evaluated in region II, cf.~\eqref{eq:mi},\eqref{eq:sol22}.  $Z_1$ is
the velocity of the discontinuity separating regions I and II, and
$Z_2$ the Velocity of the discontinuity separating regions II and III,
cf.~\eqref{eq:sol23}. (a)~Overtaking collisions
$(\Lambda_1,\Lambda_2)=(1.05,1.10)$. (b)~Head-on collisions
$(\Lambda_1,\Lambda_2)=(-1.05,1.10)$.}
\label{fig:ov2}
\end{figure*}

The averaging of the $50$ numerical solutions yields the statistical
moments of the nonlinear wave fields of the RNLS dispersive
hydrodynamics. Fig.~\ref{fig:rsol2} displays the comparison between
$\langle \rho(x,t) \rangle$ obtained numerically and the analytical
solution~\eqref{eq:mi},\eqref{eq:sol21}
for (i) and (ii).  Note that the discontinuities in
$\langle \rho(x,t)\rangle$ have a finite slope in Fig.~\ref{fig:rsol2}
which is an artifact of the averaging procedure detailed in
Appendix~\ref{sec:num}. The comparison between the numerical values of
$\langle \rho \rangle$, $Z_1$, $Z_2$ fitted from the numerical
solution $\langle \rho(x,t=5000) \rangle$, and the analytical
solutions~\eqref{eq:sol22},\eqref{eq:sol23} for different values of
$\Lambda_2$ is displayed in Fig.~\ref{fig:ov2}. The comparison shows
a good agreement between analytical and numerical solutions and
highlights the contrasting effects of (i) overtaking and (ii) head-on
collisions.  As predicted $Z_1 < C_1< C_2<Z_2$ in the case (i), whereas
$C_1 < Z_1 <Z_2 < C_2$ in the case (ii).  In the case (ii)
$\langle \rho \rangle$ in the region of interaction is almost equal to
the average value of $\rho$ for a non-interacting soliton gas
(cf. solution~\eqref{eq:sol_free}). This is due to the weakness of the
head-on interaction, clearly displayed in the comparison between
$\Delta_{+-}$ and $\Delta_{++}$ in Fig.~\ref{fig:phase_shift}.

The discrepancy between the analytical and numerical solutions can be
associated to the numerical implementation and the time-evolution of
the soliton gas.  The construction of the soliton gas at $t=0$,
detailed in Appendix~\ref{sec:imp}, is only valid if the overlap
between solitons is negligible, which is not exactly the case for the
parameters considered in Table~\ref{tab:num}. Since
$\sqrt{\Lambda_2^2-1}$ is the typical width of the
$\Lambda_2$-soliton, the overlap between solitons becomes more
important as $\Lambda_2$ decreases for a fixed initial density
$\zeta_0$. Besides the numerical scheme utilized to solve the RNLS
equation is only valid for small amplitude solitons,
cf. Appendix~\ref{sec:scheme}, and the discrepancy also increases as
$\Lambda_2$ increases.


\begin{figure}[h]
\centering
\includegraphics{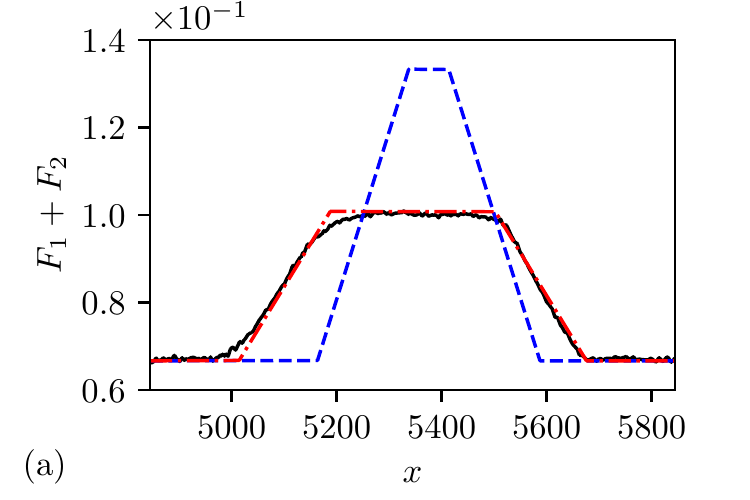}
\includegraphics{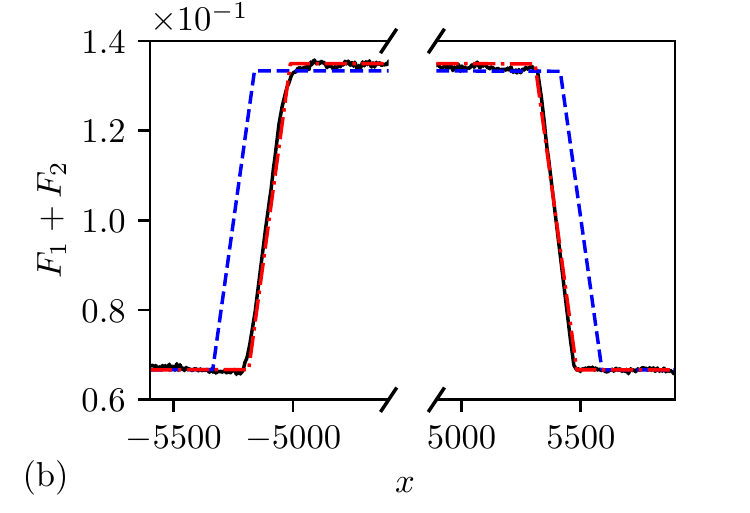}
\caption{Comparison between the total density of the soliton gas
obtained by direct numerical solution of the RNLS soliton gas shock
tube problem (black solid line) and the corresponding spectral
analytical solution $F_1(x,t)+F_2(x,t)$ where $F_i$ is given
by~\eqref{eq:sol21} (dash-dotted line); the total density in the
region of interaction $F_1^{\rm II}+F_2^{\rm II}$ is given
by~\eqref{eq:solt}. The dashed line corresponds to the total density
$F_1^{\rm free}+F_2^{\rm free}$, cf.~\eqref{eq:sol_free}. (a)~Overtaking
collisions $(\Lambda_1,\Lambda_2)=(1.05,1.10)$. (b)~Head-on collisions
$(\Lambda_1,\Lambda_2)=(-1.05,1.10)$.}
\label{fig:solt}
\end{figure}

Additionally, we can compute the variation of
the components $F_1(x,t)$ and $F_2(x,t)$ of the DOS using the
expression~\eqref{eq:mi}:
\begin{equation}
\begin{pmatrix}
F_1(x,t) \\ F_2(x,t)
\end{pmatrix} =
\begin{pmatrix}
\overline{\eta}(\Lambda_1) & \overline{\eta}(\Lambda_2)\\
\overline{u}(\Lambda_1) & \overline{u}(\Lambda_2)
\end{pmatrix}^{-1}
\begin{pmatrix}
\langle \rho (x,t) -1\rangle \\ \langle u(x,t) \rangle
\end{pmatrix},
\end{equation}
providing that the determinant
$\overline{\eta}(\Lambda_1)\overline{u}(\Lambda_2)
-\overline{\eta}(\Lambda_2) \overline{u}(\Lambda_1)$ does not vanish.
In particular, we can evaluate numerically the total density
$F_1(x,t)+F_2(x,t)$ from the numerical solutions. Fig.~\ref{fig:solt}
displays the comparison of the total density corresponding to the
examples presented in Fig.~\ref{fig:rsol2}.  Notice that, since
$\langle \rho \rangle = \overline{\eta}(\Lambda_1) F_1 +
\overline{\eta}(\Lambda_2) F_2$ with $\overline{\eta}>0$, the
variation of the moment $\langle \rho \rangle$ and the variation of
the total density $F_1+F_2$ are qualitatively similar. As expected the
RNLS soliton gas rarefies when solitons interact with overtaking
collisions ($F_1^{\rm II}+F_2^{\rm II} < 2\zeta_0$), and condenses
with head-on collisions ($F_1^{\rm II}+F_2^{\rm II} > 2\zeta_0$). As
pointed out previously, the total density in the example (ii) is very
close to the total density of the non-interacting gas $2\zeta_0$
because of the weakness of the phase shift induced by head-on
collisions.

\subsubsection{Three-component gas}
\label{sec:ex2}

We consider now the case of three-component gases with one component
belonging to the slow spectral branch and two components belonging to
the fast branch for: (iii) the DNLS equation, (iv) the RNLS
equation. Note that in the latter case the anisotropic soliton gas
features both overtaking collisions and head-on collisions. Although
one can formally solve the equation of state~\eqref{eq:S_NLS} to
obtain the expression of $S_i(\bs F)$ and solve the Rankine-Hugoniot
condition~\eqref{eq:RH}, the analytical expressions do not read as
easily as the expressions obtained in the two-component case.  We
choose here to solve~\eqref{eq:S_NLS},\eqref{eq:RH} numerically. The
values $\bs F^{\rm L}$ and $\bs F^{\rm R}$ of the initial soliton
densities considered numerically are indicated in Table~\ref{tab:num}.

\begin{figure}[h]
\includegraphics{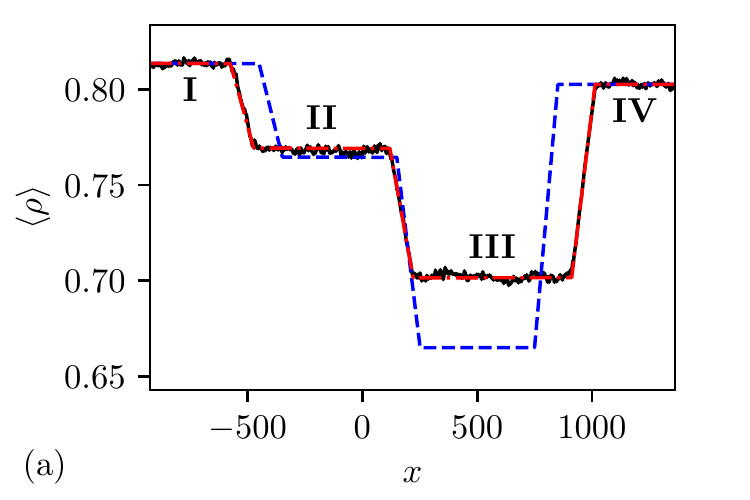}
\includegraphics{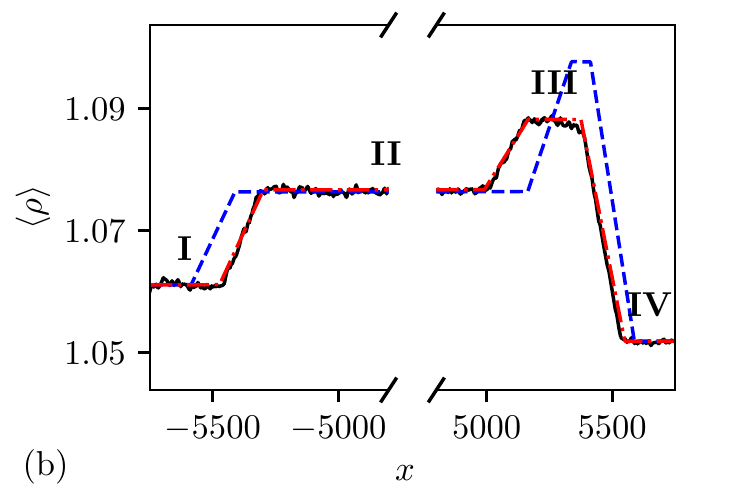}
\caption{Comparison between the ensemble average
$\langle \rho(x,t) \rangle$ obtained by direct numerical solution of
the soliton gas shock tube problem (solid line) and the analytical
solution~\eqref{eq:mi},\eqref{eq:sol} (dash-dotted line). The dashed
line corresponds to the average in the soliton gas composed of
non-interacting solitons with the spectral distribution given by
\eqref{eq:mi},\eqref{eq:sol_free}. (a)~DNLS soliton gas at
$t=2000$, case (iii). (b)~RNLS soliton gas at
$t=5000$, case (iv).}
\label{fig:psi2b}
\end{figure}

$100$ initial conditions $\rho(x,0),u(x,0)$ are realized according to
the initial spectral step distribution~\eqref{eq:init} and evolved
through a direct numerical simulation of the NLS
equation~\eqref{eq:NLS}. The statistics of the soliton gas is then
obtained by computing the average $\langle \rho (x,t)\rangle$ from the
evolution of the $100$ realizations.  Fig.~\ref{fig:psi2b} displays
the variation of the statistical moment $\langle \rho(x,t)
\rangle$. As expected, the solution is composed of 4 plateaus, where
regions II and III contain at least two distinct soliton components
and are region of interactions. The comparison in Fig.~\ref{fig:psi2b}
shows a good agreement between the analytical
solution~\eqref{eq:mi},\eqref{eq:sol_free} and the statistical
averages of the numerical solutions.

\section{Conclusions and Outlook}

In this work we have developed the spectral kinetic theory of soliton
gases in bidirectional integrable dispersive hydrodynamic
systems. Previously, such theory had been developed for (effectively)
unidirectional soliton gases, in which all pairwise soliton collisions
are characterized by a single expression for the phase shift.
Generally, however, the phase shifts in the overtaking and head-on
collisions of solitons are essentially different, which necessitates
the extension of the existing theory to bidirectional case. This
extension is also motivated by the recent experimental results on the
generation of bidirectional shallow water soliton
gases~\cite{redor_experimental_2019,redor_analysis_2020}.

The definitive quantitative characteristics of an integrable soliton
gas is the density of states (DOS), which is the density function
$f(\lambda, x, t)$ in the spectral (IST) $(x, \lambda)$-phase space.
The DOS evolution in a unidirectional non-uniform soliton gas is
governed by the kinetic equation consisting of the continuity
equation~\eqref{eq:conserv} complemented by the integral equation of
state~\eqref{eq:state} relating the soliton gas velocity and the DOS.
The presence of two distinct species of solitons corresponding to the
slow and fast branches of the dispersion relation in bidirectional
systems naturally calls for the introduction of two respective DOS's.
As a result, one arrives at a system of two coupled kinetic equations
which is the subject of the present work.

We introduced the notion of isotropic and anisotropic bidirectional
soliton gases based on the sign properties of the phase shifts in
overtaking and head-on soliton collisions in a bidirectional gas.  In
the anisotropic case, where the distinction between overtaking and
head-on soliton collisions is genuine, the kinetic of the gas is
governed by two coupled equations~\eqref{eq:kin2},\eqref{eq:s2} which
we obtained using an extension of the direct physical approach
proposed in~\cite{el_kinetic_2005}.  The approach
of~\cite{el_kinetic_2005} combines the qualitative ideas of the
original Zakharov's paper~\cite{zakharov_kinetic_1971} with the
mathematical developments of~\cite{el_thermodynamic_2003} based on the
spectral finite-gap theory. In the isotropic case, the coupled
system~\eqref{eq:kin2},\eqref{eq:s2} reduces to a single kinetic
equation~\eqref{eq:conserv},\eqref{eq:state} making a bidirectional
isotropic gas effectively equivalent to a unidirectional gas.

To highlight the principal differences between isotropic and
anisotropic soliton gases we have considered two prototypical
physically relevant examples: the (isotropic) soliton gas of the
classical defocusing NLS (DNLS) equation~\eqref{eq:DNLS} and the
(anisotropic) soliton gas of the so-called resonant NLS (RNLS)
equation~\eqref{eq:RNLS} having applications in dispersive
magneto-hydrodynamics~\cite{gurevich_origin_1988, lee_resonant_2007}.
The results for the RNLS equation are also extended to the
Kaup-Boussinesq (KB) system~\eqref{eq:KB} describing bidirectional
shallow water waves.

To provide connection between the spectral kinetics of soliton gases
and the dynamics of the physical parameters of the associated
nonlinear wave fields, we have developed a general simple procedure
enabling the evaluation of the basic ensemble averages of the soliton
gas wave field in terms of the appropriate moments of the spectral
DOS.

As an application of the developed kinetic theory we have considered
the generalized Riemann (shock-tube) problem describing collision of
several monochromatic soliton beams, each consisting of solitons
with nearly identical spectral parameters. The interaction dynamics of
such beams is described by certain exact hydrodynamic reductions of
the spectral kinetic equations.  We constructed the weak solutions of
the these hydrodynamic reductions in the form of a system of constant
states separated by propagating contact discontinuities satisfying
appropriate Rankine-Hugoniot conditions. The obtained general
solutions were then applied to the description of collisions of DNLS
and RNLS soliton gases, and the comparison with direct numerical
simulations of the DNLS and RNLS equation was made.

We stress that, although our derivation of the kinetic
equation~\eqref{eq:kin2},\eqref{eq:s2} for a dense bidirectional
soliton gas is based on the phenomenological method
of~\cite{el_kinetic_2005}, it can be formally justified using the
thermodynamic limit of the modulation equations, that has been
developed for the KdV and focusing NLS equations
in~\cite{el_thermodynamic_2003, el_spectral_2020} and can be readily
generalized to other integrable systems supporting finite-gap
solutions associated with hyperelliptic spectral Riemann
surfaces. Such a mathematical justification will be the subject of a
separate work.  Meanwhile, the excellent agreement of the exact
solutions of the Riemann problems for bidirectional kinetic equation
with appropriate direct numerical simulations for the DNLS and RNLS
equations provides a convincing confirmation of the validity of our
results.

Despite the consideration of this work being formally reliant on the
integrability of the nonlinear wave dynamics~\eqref{eq:disp_Euler},
the developed kinetic theory can be extended to non-integrable systems
supporting solitary wave solutions that exhibit nearly elastic
collisions.  An experimentally accessible example of such physical
system (albeit for a unidirectional case) is the so-called viscous
fluid conduit equation describing the dynamics of the interface
between two immiscible viscous fluids with high density and viscosity
contrast ratios, the lighter fluid being buoyantly ascending through
the heavier fluid forming a liquid ``pipe'', a
conduit~\cite{lowman_dispersive_2013}. This system supports solitary
wave solutions that exhibit nearly elastic collisions as demonstrated
numerically and confirmed
experimentally~\cite{lowman_interactions_2014}. Constructing kinetic
theory of soliton gases for non-integrable Eulerian dispersive
hydrodynamic systems of this type represents a challenging open
problem.

Another important direction of further research is the extension of
the developed kinetic theory to perturbed integrable systems.  In
particular, kinetic theory of soliton gas for the perturbed DNLS
equation could be used to describe soliton gas in a quasi-1D repulsive
BEC in a trapping potential, which has been observed experimentally
in~\cite{hamner_phase_2013}. The dynamics of the trapped condensate is
governed by the celebrated Gross-Pitaevskii equation, which is the
DNLS equation supplemented by an external potential term, which could
be treated as a perturbation in certain configurations. Although some
properties of a rarefied soliton gas in a trapped BEC have been
studied in the previous works~\cite{schmidt_non-thermal_2012,
wang_transitions_2015}, the description of a dense gas is not
available at present.  The investigation of dense soliton gas dynamics
in BECs can shed new light on turbulence in superfluids, or ``quantum
turbulence'', which has been the subject of intense research in recent
decades, see e.g.~\cite{barenghi_introduction_2014} and references
therein.

The direct experimental verification of the developed theory could be
made possible by the recent advances in the spectral synthesis of
soliton gases with a prescribed DOS~\cite{suret_nonlinear_2020}. While
the method of~\cite{suret_nonlinear_2020} was developed for deep water
waves, its extension to other types of wave propagation well described
by integrable or nearly integrable systems looks a very promising
direction since the kinetic description of soliton gases is achieved
essentially in spectral terms.
 
Concluding, we hope that our work will provide further motivation for
the theoretical and experimental study of soliton gases.

\section*{Acknowledgments}

The work was partially supported by EPSRC grant EP/R00515X/2 (T.C. and
G.E.)  and DSTL grant DSTLX-1000116851(G.E. and G.R.). The authors
thank S. Randoux, P. Suret and A. Tovbis for numerous stimulating
discussions.

\appendix

\section{Soliton solution of the KB system}
\label{sec:sol_KB}

The soliton solution~\eqref{eq:sol_RNLS} of the RNLS equation reads
after substitution in~\eqref{eq:changeKB}:
\begin{equation}
\begin{split}
\tilde \rho_{\rm s}^\pm &= 1+\lambda \tilde u_{\rm s} - \frac{\tilde
u_{\rm s}^2}{2},\\
\tilde u_{\rm s}^\pm &= \frac{2(\lambda^2-1) \left(\lambda -
\sqrt{\lambda^2-1} \tanh (\alpha/2) \right) }{2\lambda^2-1 +
\cosh(\alpha)},\\
\alpha & = \sqrt{3(\lambda^2-1)}(\tilde x-\lambda
\tilde t),
\end{split}
\end{equation}
where $\lambda^2>1$.
Note that the solution~\eqref{eq:sol_KB} is not centered at $x=0$ but
$x=\phi(\lambda)$ with
\begin{equation}
\phi(\lambda) = \frac{\sgn(\lambda)}{\sqrt{3(\lambda^2-1)}}
\log \left(|\lambda| - \sqrt{\lambda^2 -1} \right).
\end{equation}

The centered soliton solution reads:
\begin{equation}
\label{eq:sol_KB}
\begin{split}
\tilde \rho_{\rm s}^\pm(\tilde x + \phi(\lambda), \tilde t) &=
1+\frac{2(\lambda^2 -1) \big(1 + |\lambda| \cosh
(\alpha) \big)}{\big(|\lambda|
+ \cosh (\alpha) \big)^2},\\
\tilde u_{\rm s}^\pm(\tilde x + \phi(\lambda), \tilde t) &= \frac{2 \sgn(\lambda)
(\lambda^2-1)}{|\lambda|+ \cosh (\alpha)},\\
\alpha &=\sqrt{3(\lambda^2 -1 )} (\tilde x-\lambda \tilde t),
\end{split}
\end{equation}
which coincides with the solution derived
in~\cite{zhang_bidirectional_2003}.

\section{Numerical implementation of soliton gases for the
NLS equation}
\label{sec:num}

\subsection{Implementation of the step distribution}
\label{sec:imp}

We implement the soliton gas using the method developed
in~\cite{carbone_macroscopic_2016}. The initial step distribution of
the spectral Riemann problem~\eqref{eq:init} with values given in
Table~\ref{tab:num} describes a rarefied gas where solitons do not
overlap. Such a distribution is implemented by the superposition of
solitons
\begin{equation}
\label{eq:implementation}
\begin{split}
\rho(x,t=0) = \sum_i \rho_{\rm s}(x-\xi_i;\Lambda_i),\\
u(x,t=0) = \sum_i u_{\rm s}(x-\xi_i;\Lambda_i),
\end{split}
\end{equation}
where the $\Lambda_i$'s are the spectral parameters of the solitons and
the $\xi_i$'s their initial position. Although the particles' position
$\xi_i$ of an ``ideal'' soliton gas should be distributed according to
a Poisson process~\cite{el_soliton_2001}, this cannot be implemented
numerically since the solitons are not allowed to overlap.  In our
numerics, the distance between two solitons $\xi_{i+1}-\xi_i$ is
uniformly distributed in the interval $[d_1,d_2]$ with $0<d_1<d_2$
such that the solitons do not overlap; the total density of solitons
is given by $2/(d_1+d_2)$.  We choose $(d_1,d_2)=(10,20)$ for the
RNLS-Riemann problems (i), (ii) and (iv) and $(d_1,d_2)=(8,12)$ for
the DNLS-Riemann problem, cf. Sec.~\eqref{sec:riem}.

\subsection{Numerical scheme}
\label{sec:scheme}

The DNLS equation~$i \psi_t + \tfrac12 \psi_{xx}-|\psi|^2 \psi=0$,
$\psi = \sqrt \rho \exp(i\int u \,dx)$ is solved with periodic
boundary conditions $\psi(x=L,t) = \psi(x=0,t)$ using a Fourier
spectral method. The linear part of the DNLS equation is resolved with
an integrating factor and the problem is integrated in time using
fourth-order Runge-Kutta method.

Since the dispersive term in the RNLS equation~\eqref{eq:RNLS} is a
nonlinear term in $\psi$, the RNLS equation is first transformed into
the KB equation~\eqref{eq:KB} using the change of
variables~\eqref{eq:changeKB}.  The KB system is then solved with
periodic boundary conditions $\psi(x=L,t) = \psi(x=0,t)$ using a
Fourier spectral method (with a fourth-order Runge-Kutta method for
the time integration).  Eq.~\eqref{eq:KB} displays a short wavelength
instability: the amplitude of modes
$\tilde \rho-1 \propto \tilde u \propto \cos( k_i \tilde x )$ grows
exponentially with time for $k_i > \sqrt 3$.  We thus filter out
Fourier modes $k_i > \sqrt 3$ after each time step. This imposes a
constraint on the type of solitons that can be implemented
numerically. Indeed, large amplitude solitons $|\lambda| \gg 1$
populate the short-wavelength Fourier modes $k_i > \sqrt 3$ which are
not taken into account in the numerical scheme.  We thus consider in
the numerical simulations the solitons for which
$|\lambda| \in (1,1.1)$.

\subsection{Spatial and ensemble averages}

The statistical moment $\langle \rho \rangle$ determined numerically
in Sec.~\ref{sec:riem} is obtained with: 1) the average over ensemble
of $50$ or $100$ realizations and 2) a local spatial average over
the mesoscopic space interval~$\ell$ (cf.~\eqref{eq:ell}):
\begin{equation}
\ell = \frac{10}{\max\left[ \sum_i F_i(x,t=0) \right]}.
\end{equation}
As pointed out in Sec.~\ref{sec:mean}, both averaging procedures are
equivalent providing that the soliton gas is locally ergodic.  The
choice of the value for $\ell$ ensures that the space interval
contains at least $10$ solitons. Note that the transitions of the
numerically evaluated mean field field $\langle \rho(x,t) \rangle$
corresponding to contact discontinuities in the analytical solution
have a finite slope proportional to $1/\ell$ because of the spatial
averaging.


\begin{thebibliography}{10}

\bibitem{biondini_dispersive_2016} G.~Biondini, G.~El, M.~Hoefer, and
P.~Miller, ``Dispersive hydrodynamics: Preface,'' {\em Physica D:
Nonlinear Phenomena}, vol.~333, pp.~1--5, 2016.

\bibitem{maiden_solitonic_2018} M.~D. Maiden, D.~V. Anderson,
N.~A. Franco, G.~A. El, and M.~A. Hoefer, ``Solitonic dispersive
hydrodynamics: Theory and observation,'' {\em Physical Review
Letters}, vol.~120, p.~144101, 2018.

\bibitem{sprenger_hydrodynamic_2018} P.~Sprenger, M.~A. Hoefer, and
G.~A. El, ``Hydrodynamic optical soliton tunneling,'' {\em Physical
Review E}, vol.~97, p.~032218, 2018.

\bibitem{el_expansion_2016} G.~A. El, M.~A. Hoefer, and M.~Shearer,
``Expansion shock waves in regularized shallow-water theory,'' {\em
Proceedings of the Royal Society A: Mathematical, Physical and
Engineering Sciences}, vol.~472, p.~20160141, 2016.

\bibitem{el_dispersive_2016} G.~A. El and M.~A. Hoefer, ``Dispersive
shock waves and modulation theory,'' {\em Physica D: Nonlinear
Phenomena}, vol.~333, pp.~11--65, 2016.

\bibitem{zakharov_turbulence_2009} V.~E. Zakharov, ``Turbulence in
integrable systems,'' {\em Studies in Applied Mathematics}, vol.~122,
pp.~219--234, 2009.

\bibitem{gurevich_development_1999} A.~V. Gurevich, K.~P. Zybkin, and
G.~A. {\'E}l{\rq}, ``Development of stochastic oscillations in a
one-dimensional dynamical system described by the {Korteweg-de Vries}
equation,'' {\em Journal of Experimental and Theoretical Physics},
vol.~88, pp.~182--195, 1999.

\bibitem{el_dam_2016} G.~A. El, E.~G. Khamis, and A.~Tovbis, ``Dam
break problem for the focusing nonlinear {Schr{\"o}dinger} equation
and the generation of rogue waves,'' {\em Nonlinearity}, vol.~29,
pp.~2798--2836, 2016.

\bibitem{Walczak:15} P.~Walczak, S.~Randoux, and P.~Suret, ``Optical
rogue waves in integrable turbulence,'' {\em Physical Review Letters},
vol.~114, p.~143903, 2015.

\bibitem{Randoux:16b} S.~Randoux, P.~Walczak, M.~Onorato, and
P.~Suret, ``Nonlinear random optical waves: Integrable turbulence,
rogue waves and intermittency,'' {\em Physica D: Nonlinear Phenomena},
vol.~333, pp.~323--335, 2016.

\bibitem{gelash_strongly_2018} A.~A. Gelash and D.~S. Agafontsev,
``Strongly interacting soliton gas and formation of rogue waves,''
{\em Physical Review E}, vol.~98, pp.~042210--1--042210--12, 2018.

\bibitem{costa_soliton_2014} A.~Costa, A.~R. Osborne, D.~T. Resio,
S.~Alessio, E.~Chriv\`{\i}, E.~Saggese, K.~Bellomo, and C.~E. Long,
``Soliton {Turbulence} in {Shallow} {Water} {Ocean} {Surface}
{Waves},'' {\em Physical Review Letters}, vol.~113, p.~108501, 2014.

\bibitem{trillo_experimental_2016} S.~Trillo, G.~Deng, G.~Biondini,
M.~Klein, G.~F. Clauss, A.~Chabchoub, and M.~Onorato, ``Experimental
observation and theoretical description of multisoliton fission in
shallow water,'' {\em Physical Review Letters}, vol.~117, p.~144102,
2016.

\bibitem{trillo_observation_2016} S.~Trillo, M.~Klein, G.~Clauss, and
M.~Onorato, ``Observation of dispersive shock waves developing from
initial depressions in shallow water,'' {\em Physica D: Nonlinear
Phenomena}, vol.~333, pp.~276--284, 2016.

\bibitem{gelash_bound_2019} A.~Gelash, D.~Agafontsev, V.~Zakharov,
G.~El, S.~Randoux, and P.~Suret, ``Bound state soliton gas dynamics
underlying the spontaneous modulational instability,'' {\em Physical
Review Letters}, vol.~123, p.~234102, 2019.

\bibitem{el_spectral_2020} G.~El and A.~Tovbis, ``Spectral theory of
soliton and breather gases for the focusing nonlinear
{Schr{\"o}dinger} equation,'' {\em Physical Review E}, vol.~101,
p.~052207, 2020.

\bibitem{el_kinetic_2005} G.~A. El and A.~M. Kamchatnov, ``Kinetic
equation for a dense soliton gas,'' {\em Physical Review Letters},
vol.~95, p.~204101, 2005.

\bibitem{suret_nonlinear_2020} P.~Suret, A.~Tikan, F.~Bonnefoy,
F.~Copie, G.~Ducrozet, A.~Gelash, G.~Prabhudesai, G.~Michel,
A.~Cazaubiel, E.~Falcon, G.~El, and S.~Randoux, ``Nonlinear spectral
synthesis of soliton gas in deep-water surface gravity waves,''
arXiv:2006.16778 [nlin.PS], 2020.

\bibitem{el_kinetic_2011} G.~A. El, A.~M. Kamchatnov, M.~V. Pavlov,
and S.~A. Zykov, ``Kinetic {Equation} for a {Soliton} {Gas} and {Its}
{Hydrodynamic} {Reductions},'' {\em Journal of Nonlinear Science},
vol.~21, pp.~151--191, 2011.

\bibitem{ablowitz_note_1982} M.~J. Ablowitz and Y.~Kodama, ``Note on
asymptotic solutions of the {Korteweg-de Vries} equation with
solitons,'' {\em Studies in Applied Mathematics}, vol.~66,
pp.~159--170, 1982.

\bibitem{zakharov_kinetic_1971} V.~E. Zakharov, ``Kinetic equation for
solitons,'' {\em Journal of Experimental and Theoretical Physics},
vol.~33, pp.~538--541, 1971.

\bibitem{el_thermodynamic_2003} G.~A. El, ``The thermodynamic limit of
the {Whitham} equations,'' {\em Physics Letters A}, vol.~311,
pp.~374--383, 2003.

\bibitem{el_critical_2016} G.~A. El, ``Critical density of a soliton
gas,'' {\em Chaos: An Interdisciplinary Journal of Nonlinear Science},
vol.~26, p.~023105, 2016.

\bibitem{doyon_soliton_2018} B.~Doyon, T.~Yoshimura, and J.-S. Caux,
``Soliton {Gases} and {Generalized} {Hydrodynamics},'' {\em Physical
Review Letters}, vol.~120, p.~045301, 2018.

\bibitem{doyon_geometric_2018} B.~Doyon, H.~Spohn, and T.~Yoshimura,
``A geometric viewpoint on generalized hydrodynamics,'' {\em Nuclear
Physics B}, vol.~926, pp.~570--583, 2018.

\bibitem{vu_equations_2019} D.-L. Vu and T.~Yoshimura, ``Equations of
state in generalized hydrodynamics,'' {\em SciPost Physics}, vol.~6,
p.~23, 2019.

\bibitem{whitham_linear_1999} G.~B. Whitham, {\em Linear and Nonlinear
Waves}.  \newblock John Wiley \& Sons, Inc., 1999.

\bibitem{kaup_higher_1975} D.~J. Kaup, ``A higher-order water-wave
equation and the method for solving it,'' {\em Progress of Theoretical
Physics}, vol.~54, pp.~396--408, 1975.

\bibitem{kamchatnov_nonlinear_2000} A.~M. Kamchatnov, {\em Nonlinear
periodic waves and their modulations: an introductory course}.
\newblock World Scientific, 2000.

\bibitem{abanov_integrable_2009} A.~G. Abanov, E.~Bettelheim, and
P.~Wiegmann, ``Integrable hydrodynamics of calogero--sutherland model:
bidirectional benjamin--ono equation,'' {\em Journal of Physics A:
Mathematical and Theoretical}, vol.~42, p.~135201, 2009.

\bibitem{redor_experimental_2019} I.~Redor, E.~Barth{\'e}lemy,
H.~Michallet, M.~Onorato, and N.~Mordant, ``Experimental {Evidence} of
a {Hydrodynamic} {Soliton} {Gas},'' {\em Physical Review Letters},
vol.~122, no.~21, p.~214502, 2019.

\bibitem{redor_analysis_2020} I.~Redor, E.~Barth{\'e}lemy, N.~Mordant,
and H.~Michallet, ``Analysis of soliton gas with large-scale
video-based wave measurements,'' {\em Experiments in Fluids}, vol.~61,
p.~216, 2020.

\bibitem{li_bidirectional_2003} Y.~Li and J.~E. Zhang, ``Bidirectional
soliton solutions of the classical {Boussinesq} system and {AKNS}
system,'' {\em Chaos, Solitons \& Fractals}, vol.~16, pp.~271--277,
2003.

\bibitem{zakharov_interaction_1973} V.~E. Zakharov and A.~B. Shabat,
``Interaction between solitons in a stable medium,'' {\em Journal of
Experimental and Theoretical Physics}, vol.~37, pp.~823--828, 1973.

\bibitem{lee_resonant_2007} J.-H. Lee, O.~Pashaev, C.~Rogers, and
W.~Schief, ``The resonant nonlinear {Schr{\"o}dinger} equation in cold
plasma physics. application of {B{\"a}cklund}--{Darboux}
transformations and superposition principles,'' {\em Journal of Plasma
Physics}, vol.~73, pp.~257--272, 2007.

\bibitem{yang_nonlinear_2010} J.~Yang, {\em Nonlinear Waves in
Integrable and Nonintegrable Systems}.  \newblock Society for
Industrial and Applied Mathematics, 2010.

\bibitem{schmidt_non-thermal_2012} M.~Schmidt, S.~Erne, B.~Nowak,
D.~Sexty, and T.~Gasenzer, ``Non-thermal fixed points and solitons in
a one-dimensional Bose gas,'' {\em New Journal of Physics}, vol.~14,
p.~075005, 2012.

\bibitem{wang_transitions_2015} W.~Wang and P.~G. Kevrekidis,
``Transitions from order to disorder in multiple dark and multiple
dark-bright soliton atomic clouds,'' {\em Physical Review E}, vol.~91,
p.~032905, 2015.

\bibitem{lee_solitons_2007} J.-H. Lee and O.~K. Pashaev, ``Solitons of
the resonant nonlinear {Schr{\"o}dinger} equation with nontrivial
boundary conditions: Hirota bilinear method,'' {\em Theoretical and
Mathematical Physics}, vol.~152, pp.~991--1003, 2007.

\bibitem{gurevich_origin_1988} A.~V. Gurevich and A.~L. Krylov, ``The
origin of a nondissipative shock wave,'' {\em Doklady Physics: A
Journal of the Russian Academy of Sciences}, vol.~33, no.~8,
pp.~603--605, 1988.

\bibitem{el_soliton_2001} G.~A. El, A.~L. Krylov, S.~Molchanov, and
S.~Venakides, ``Soliton turbulence as a thermodynamic limit of
stochastic soliton lattices,'' {\em Physica D: Nonlinear Phenomena},
vol.~152-153, pp.~653--664, 2001.

\bibitem{flaschka_multiphase_1980} H.~Flaschka, M.~G. Forest, and
D.~W. McLaughlin, ``Multiphase averaging and the inverse spectral
solution of the {Korteweg-de Vries} equation,'' {\em Communications on
Pure and Applied Mathematics}, vol.~33, pp.~739--784, 1980.

\bibitem{moser_integrable_1981} J.~Moser, {\em Integrable Hamiltonian
Systems and Spectral Theory}.  \newblock Scuola normale superiore,
1981.

\bibitem{shurgalina_nonlinear_2016} E.~G. Shurgalina and
E.~N. Pelinovsky, ``Nonlinear dynamics of a soliton gas: {Modified
Korteweg--de Vries} equation framework,'' {\em Physics Letters A},
vol.~380, pp.~2049--2053, 2016.

\bibitem{dutykh_numerical_2014} D.~Dutykh and E.~Pelinovsky,
``Numerical simulation of a solitonic gas in {KdV} and {KdV}--{BBM}
equations,'' {\em Physics Letters A}, vol.~378, pp.~3102--3110, 2014.

\bibitem{pavlov_generalized_2012} M.~V. Pavlov, V.~B. Taranov, and
G.~A. El, ``Generalized hydrodynamic reductions of the kinetic
equation for a soliton gas,'' {\em Theoretical and Mathematical
Physics}, vol.~171, pp.~675--682, 2012.

\bibitem{carbone_macroscopic_2016} F.~Carbone, D.~Dutykh, and
G.~A. El, ``Macroscopic dynamics of incoherent soliton ensembles:
{Soliton} gas kinetics and direct numerical modelling,'' {\em EPL
(Europhysics Letters)}, vol.~113, p.~30003, 2016.

\bibitem{ferapontov_integration_1991} E.~Ferapontov, ``Integration of
weakly nonlinear hydrodynamic systems in {Riemann} invariats,'' {\em
Physics Letters A}, vol.~158, pp.~112--118, 1991.

\bibitem{tsarev_geometry_1991} S.~P. Tsar{\"e}v, ``The geometry of
hamiltonian systems of hydrodynamic type.  the generalized hodograph
method,'' {\em Mathematics of the USSR-Izvestiya}, vol.~37,
pp.~397--419, 1991.

\bibitem{sprenger_discontinuous_2020} P.~Sprenger and M.~A. Hoefer,
``Discontinuous shock solutions of the whitham modulation equations as
zero dispersion limits of traveling waves,'' {\em Nonlinearity},
vol.~33, pp.~3268--3302, 2020.

\bibitem{gavrilyuk_stationary_2020} S.~Gavrilyuk, B.~Nkonga,
K.-M. Shyue, and L.~Truskinovsky, ``Stationary shock-like transition
fronts in dispersive systems,'' {\em Nonlinearity}, vol.~33,
pp.~5477--5509, 2020.

\bibitem{castro-alvaredo_emergent_2016} O.~A. Castro-Alvaredo, B.~Doyon, and
T.~Yoshimura, ``Emergent hydrodynamics in integrable quantum systems
out of equilibrium,'' {\em Physical Review X}, vol.~6, p.~041065,
2016.

\bibitem{bertini_transport_2016} B.~Bertini, M.~Collura, J.~De Nardis,
and M. Fagotti, ``Transport in out-of-qquilibrium XXZ chains: exact
profiles of charges and currents,'', {\em Physical Review Letters},
vol.~117, p.~207201, 2016.

\bibitem{doyon_dynamics_2017} B.~Doyon and H.~Spohn, ``Dynamics of
hard rods with initial domain wall state,'' {\em Journal of
Statistical Mechanics:Theory and Experiment}, vol.~7, p.~073210, 2017.

\bibitem{kuniba_generalized_2020} A.~Kuniba, G.~Misguich, and
V.~Pasquier, ``Generalized hydrodynamics in box-ball system,'' {\em
Journal of Physics A: Mathematical and Theoretical}, vol.~53,
p.~404001, 2020.

\bibitem{croydon_generalized_2020} D.~A. Croydon and M.~Sasada,
``Generalized Hydrodynamic Limit for the Box–Ball System,'' {\em
Communications in Mathematical Physics}, 2020.

\bibitem{rozhdestvenskii_systems_1983} B.~Rozhdestvenskii and
N.~Janenko, {\em Systems of quasilinear equations and their
applications to gas dynamics}.  \newblock Providence: RI: American
Mathematical Society, 1983.

\bibitem{lax_hyperbolic_1973} P.~D. Lax, {\em Hyperbolic systems of
conservation laws and the mathematical theory of shock waves}.
\newblock Society for Industrial and Applied Mathematics, 1973.

\bibitem{hamner_phase_2013} C.~Hamner, Y.~Zhang, J.~J. Chang,
C.~Zhang, and P.~Engels, ``Phase winding a two-component Bose-Einstein
Condensate in an elongated trap: experimental observation of moving
magnetic orders and dark-bright solitons,'' {\em Physical Review
Letters}, vol.~111, p.~264101, 2013.

\bibitem{barenghi_introduction_2014} C.~F. Barenghi, L.~Skrbek, and
K.~R. Sreenivasan, ``Introduction to quantum turbulence,'' {\em
Proceedings of the National Academy of Sciences}, vol.~111,
pp.~4647--4652, 2014.

\bibitem{lowman_dispersive_2013} N.~K. Lowman and M.~A. Hoefer,
``Dispersive hydrodynamics in viscous fluid conduits,'' {\em Physical
Review E}, vol.~88, p.~023016, 2013.

\bibitem{lowman_interactions_2014} N.~K. Lowman, M.~A. Hoefer, and
G.~A. El, ``Interactions of large amplitude solitary waves in viscous
fluid conduits,'' {\em Journal of Fluid Mechanics}, vol.~750,
pp.~372--384, 2014.

\bibitem{zhang_bidirectional_2003} J.~E. Zhang and Y.~Li,
``Bidirectional solitons on water,'' {\em Physical Review E}, vol.~67,
p.~016306, 2003.

\end{thebibliography}
\end{document}